\documentclass[english,twocolumn]{revtex4}
\usepackage[L7x,T1]{fontenc}
\usepackage[latin9]{inputenc}
\usepackage{xcolor}
\usepackage{pdfcolmk}
\usepackage{array}
\usepackage{bm}
\usepackage{multirow}
\usepackage{amsmath}
\usepackage{amssymb}
\usepackage{graphicx}
\usepackage{esint}
\usepackage{bm}

\makeatother

\usepackage{babel}
\begin{document}
\global\long\def\deriv#1#2{\frac{\mathrm{d}#1}{\mathrm{d}#2}}
\global\long\def\pderiv#1#2{\frac{\partial#1}{\partial#2}}
\global\long\def\emath{\mathrm{e}}
\global\long\def\d{\mathrm{d}}
\global\long\def\vec#1{\bm{#1}}
\global\long\def\icm{\mathrm{\mathrm{cm}^{-1}}}
\global\long\def\imath{\mathrm{i}}
\global\long\def\ptderiv{\frac{\partial}{\partial t}}
\renewcommand{\figurename}{Fig.}

\title{Vibronic phenomena and exciton--vibrational interference in two-dimensional
spectra of molecular aggregates}

\author{Vytautas Butkus}

\affiliation{Department of Theoretical Physics, Faculty of Physics, Vilnius University,
Sauletekio 9-III, 10222 Vilnius, Lithuania}

\affiliation{Center for Physical Sciences and Technology, Gostauto 9, 01108 Vilnius,
Lithuania}

\author{Leonas Valkunas}

\affiliation{Department of Theoretical Physics, Faculty of Physics, Vilnius University,
Sauletekio 9-III, 10222 Vilnius, Lithuania}

\affiliation{Center for Physical Sciences and Technology, Gostauto 9, 01108 Vilnius,
Lithuania}

\author{Darius Abramavicius}

\email{darius.abramavicius@ff.vu.lt}

\affiliation{Department of Theoretical Physics, Faculty of Physics, Vilnius University,
Sauletekio 9-III, 10222 Vilnius, Lithuania}

\affiliation{State Key Laboratory of Supramolecular Complexes, Jilin University,
2699 Qianjin Street, Changchun 130012, PR China}
\begin{abstract}
A general theory of electronic excitations in aggregates of molecules
coupled to intramolecular vibrations and the harmonic environment
is developed for simulation of the third-order nonlinear spectroscopy
signals. The model is applied in studies of the time-resolved two-dimensional
coherent spectra of four characteristic model systems: weakly / strongly
vibronically coupled molecular dimers coupled to high / low frequency
intramolecular vibrations. The results allow us to classify the typical
spectroscopic features as well as to define the limiting cases, when
the long-lived quantum coherences are present due to vibrational lifetime
borrowing, when the complete exciton-vibronic mixing occurs and when
separation of excitonic and vibrational coherences is proper.
\end{abstract}
\maketitle
\clearpage{}

\section{Introduction}

Excitonic energy spectrum of molecular aggregates experiences essential
transformation due to the presence of high-frequency intramolecular
vibrations. As a result, coupling between electronic excitations and
intramolecular vibrations known as vibronic coupling turn to be responsible
for a host of spectroscopically-observed phenomena. The vibronic effects
have been investigated intensively by different theoretical methods
since the foundation of molecular (Frenkel) exciton theory \cite{Davydov-book,V.I.Broude1985}.
Along with the advance of nonlinear spectroscopic techniques, some
new insights related to coupling between electronic degrees of freedom
of molecular aggregates and intramolecular vibrations were observed
in third-order spectroscopic signals, for example, in two-dimensional
(2D) coherent spectra demonstrating vibrational wave-packet motion,
long-lived coherences, vibrational anisotropy beats, polaron formation,
etc.\cite{nemeth-sperling-JCP2010,Egorova2007,Smith2011,Dahlbom2002,Gelzinis2011,ZhaoYang_molecular_ring_JCP2013}.
Probably the most extensively discussed issue lately is the impact
of discrete vibrational resonances on the electronic coherences and
\emph{vice versa.} These coherences are observed in the 2D electronic
spectroscopy, but its possible role in energy transfer is currently
under discussion \cite{Chin2013,Christensson_JPCB2012,Kreisbeck2012}.

Range and diversity of molecular systems, where vibronic coupling
is very significant, appears to be extremely wide. Historically, molecular
crystals were the first systems where the vibronic coupling was considered
and the theoretical basis of the spectral characterization was developed
by analyzing their stationary spectra \cite{Fulton1964,Philpott1971,Davydov1970,Davydov1971a}.
Further development of the theoretical approach was addressed to studies
of vibronic excitations in H and J aggregates and in molecular films
\cite{Friesner1981,Scherer1984}. Strong coupling to discrete intramolecular
high-frequency modes of the ${\rm C=C}$ stretch vibration at around
$1400\,\icm$ together with the strong electrostatic interaction between
the molecules are the most evident properties of the J-aggregates.
Coupling to discrete low-frequency intramolecular modes (160~$\icm$,
for example\cite{Milota2013_JPCA_VibrJaggr,KobayashiBook1996}) has
also been considered. Significant vibronic features are prevalent
in spectra of aggregated and strongly-coupled molecular dimeric dyes,
the formation of which is usually the first step towards the large-scale
molecular aggregation \cite{West1965,Kopainsky1981,Baraldi2002,Moran2006,Seibt2008,Bixner2012}. 

Photosynthetic pigment--protein (P--P) complexes could be considered
as yet another class of molecular systems, where weak vibronic coupling
(however, only recently observed) was found to be important \cite{Kolli2012,Christensson_JPCB2012,Adolphs2006,Lee-Fleming2007,Womick2011,Richards2012}.
Since pigment molecules within P--P formations are weakly-coupled
and the surrounding protein framework is ready to dissipate any vibrational
motion of the pigments, the domination of electronic coupling over
vibronic effects is commonly assumed. Therefore, long-lasting oscillations
in coherent 2D spectra were initially explained by purely excitonic
coherences \cite{engel-nat2007,ColliniScholes2010,Panitchayangkoon2011}.
Recently, vibronic components and mixing of both, electronic and vibronic,
ingredients have been reported \cite{Christensson2011,Jonas_PNAS2012}.

\begin{figure}
\includegraphics{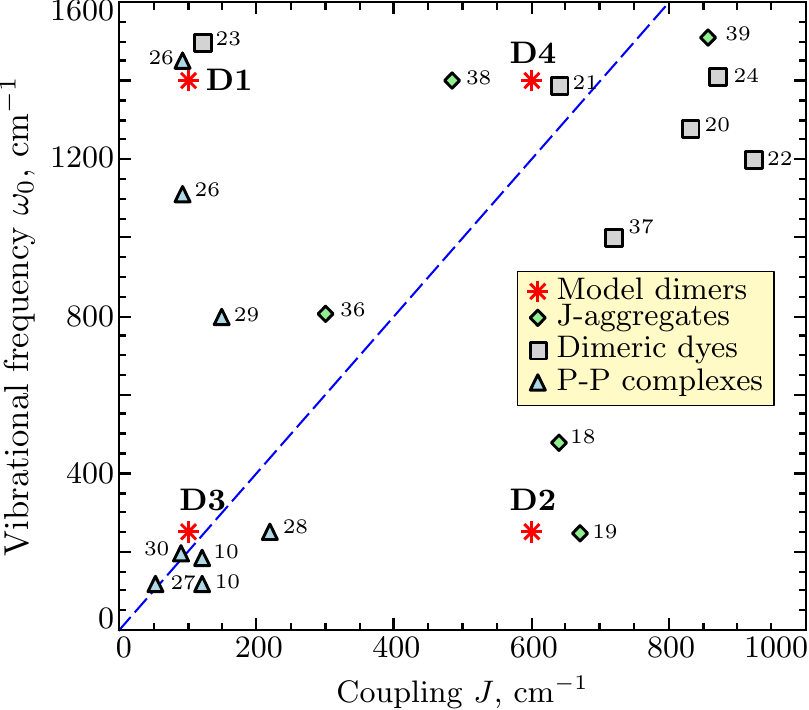}

\protect\caption{\label{fig:(a)-Experimentally-and}Experimentally and theoretically
investigated molecular systems (dimeric dyes, weakly-coupled P-P complexes,
J-aggregates and films), characterized by different electronic resonance
interactions $J$ and vibrational frequencies $\omega_{0}$. Dashed
line indicates the region of exciton-vibronic resonance ($\omega_{0}=2J$),
the numbers next to symbols are the references to the corresponding
studies. Stars indicate the model dimer systems considered in this
paper. }
\end{figure}

If we were to represent the above-mentioned systems as points on a
schematic two-dimensional phase space, where the axes indicate vibrational
frequency and electronic resonance interaction, the most of it would
be covered as presented in Fig.~\ref{fig:(a)-Experimentally-and}.
We can make a classification of the points scattered over the plot
by considering the possible time-resolved experiment with ultra-short
laser pulses of typical bandwidth of $\sim1000-2000$ $\icm$. In
the top--left corner of the figure we then have the weakly-coupled
systems with high-frequency vibrations. In this case the experiment
would resolve a few peaks of the vibronic progression at most and
the splittings due to electronic coupling would be overlapping. The
mixed case where electronic resonance interactions and vibronic progression
would be resolvable along with strong quantum-mechanical mixing of
both types of transitions, is present in the top-right part of the
scheme. The laser spectrum would cover only a few peaks in this case.
On the bottom--left corner we have the mixed systems again, but the
laser pulse spectrum could cover all peaks. And, finally, on the bottom--right
corner we have the case where the full vibrational progression could
be observed in the experiment and the electronic splitting would be
well-resolved.

To cover all these cases in an unified model, we present the molecular
exciton--vibronic theory developed for the purpose of its application
in describing the two-dimensional electronic spectroscopy signals.
It is based on the Holstein-type exciton--vibronic Hamiltonian with
assumption of multi-particle vibronic state basis \cite{Holstein1959,Fulton1964}.
Dissipation is included by coupling the vibrational coordinate to
the harmonic bath. Special attention is paid to exciton--vibronic
resonances, at which the most pronounced mixing of states is present
\cite{Polyutov2012,Chenu2013,Jonas_PNAS2012,Butkus2013}. Four different
models of dimer systems are chosen for consideration as indicated
by stars in Fig.~\ref{fig:(a)-Experimentally-and}: two being considerably
away from the exciton--vibronic resonance (\textbf{D1 }and \textbf{D2})
and another two corresponding to mixed conditions (\textbf{D3} and
\textbf{D4}). As one can observe, these models represent four typical
molecular systems: weakly-coupled P--P complexes with high and low-frequency
vibrations (\textbf{D1} and \textbf{D3}), the J-aggregate (\textbf{D2})
and a molecular dye (\textbf{D4}). Therefore, the conclusions drawn
from the results of model systems are general in terms of its application
to different molecular aggregates.

\section{\label{sub:ELECTRONIC_DIMER-1}Vibrational aggregate model}

Various models of a molecule coupled to continuum of bath vibrations
were developed within the framework of the perturbative system--bath
interaction expansion\cite{mukbook}. The bath is then described by
the spectral density function, which represents auto-correlations
of the electronic \emph{site} energy fluctuations due to the environment.
The most popular model assumes the Brownian particle-like vibrational
motion of the molecule in a solvent \cite{May2011,Valkunas2013}.
This model is usually enough to obtain proper spectral lineshapes
in simulations of systems with no expressed high-frequency vibrations
at fixed temperature. For systems with well-resolved high-frequency
modes of vibrations the spectral density approach is applied by including
a $\delta$-shaped or finite-bandwidth peak into the bath spectral
density function. The $\delta$-peak does give ever-lasting coherent
beats in the coherent 2D spectra\cite{mancal-sperling-jcp2010,Egorova2008},
while in case of finite-width peak decay of oscillations is obtained
due to pure dephasing \cite{Butkus-Abramavicius-Valkunas-JCP2012,Seibt2013}.
However, this method has two deficiencies. Firstly, it neglects the
effects caused by quantum-mechanical mixing of the vibronic levels
of different molecules when the vibronic splitting is comparable to
the intermolecular excitonic coupling. Secondly, it does not include
vibrational relaxation as the vibrations are assumed to be in thermal
equilibrium at fixed temperature. These could be important effects
when the coupling to vibrations is strong.

There have been several studies of nonlinear coherent spectra of molecular
dimers with exciton--vibronic mixing included \cite{Chenu2013,Jonas_PNAS2012,Krcmar2013_CP}.
However, the realistic molecular aggregates contain several dozens
or hundreds of molecules. We develop a general description applicable
for molecular aggregates with an arbitrary number of chromophores.

Let us start with the displaced oscillator model of a molecule. It
dictates that the Hamiltonian of a single (say $m$-th) molecule in
an aggregate can be given by 
\begin{align}
\hat{H}_{m} & =\left[\frac{\hat{p}_{m}^{2}}{2}+\frac{\omega_{m}^{2}}{2}\hat{q}_{m}^{2}\right]|{\rm g}^{m}\rangle\langle{\rm g}^{m}|\nonumber \\
 & +\left[\epsilon_{m}+\frac{\hat{p}_{m}^{2}}{2}+\frac{\omega_{m}^{2}}{2}(\hat{q}_{m}-d_{m})^{2}\right]|{\rm e}^{m}\rangle\langle{\rm e}^{m}|.
\end{align}
Here $\hat{p}_{m}$ and $\hat{q}_{m}$ are the momentum and coordinate
operators of the intramolecular vibrational motion, $\omega_{m}$
is the vibrational frequency and $d_{m}$ is the displacement in the
excited state. The effective mass of the oscillator is taken as unity.
Ground and excited state wavevectors for the $m$-th molecule, $|{\rm g}^{m}\rangle$
and $|{\rm e}^{m}\rangle$ (we use superscript indices for later convenience)
respectively, in the space of electronic states of the single molecule
comprise the complete basis set, thus $|{\rm g}^{m}\rangle\langle{\rm g}^{m}|+|{\rm e}^{m}\rangle\langle{\rm e}^{m}|=1$.
After introducing operators for electronic excitations $\hat{B}_{m}^{\dagger}$
, so that $|{\rm e}^{m}\rangle=\hat{B}_{m}^{\dagger}|{\rm g}^{m}\rangle$,
and its Hermitian conjugate $\hat{B}_{m}$, we can write 
\begin{align}
\hat{H}_{m} & =\frac{\hat{p}_{m}^{2}}{2}+\frac{\omega_{m}^{2}}{2}\hat{q}_{m}^{2}\nonumber \\
 & +\left(\epsilon_{m}+\lambda_{m}-\omega_{m}^{2}d_{m}\hat{q}_{m}\right)\hat{B}_{m}^{\dagger}\hat{B}_{m}.\label{eq:Ham-El}
\end{align}
Here we defined the reorganization energy $\lambda_{m}=\omega_{m}^{2}d_{m}^{2}/2$.
As the molecule can be electronically excited just once, we must have
$\hat{B}_{m}^{\dagger}|{\rm e}^{m}\rangle=0$ or $\hat{B}_{m}\hat{B}_{m}^{\dagger}+\hat{B}_{m}^{\dagger}\hat{B}_{m}=1,$
which reflects the fermionic property. 

By inserting the bosonic creation and annihilation operators for the
vibrational degrees of freedom 
\begin{align*}
\hat{p}_{m} & ={\rm i}\sqrt{\frac{\omega_{m}}{2}}\left(\hat{b}_{m}^{\dagger}-\hat{b}_{m}\right)~{\rm and}~\hat{q}_{m}=\sqrt{\frac{1}{2\omega_{m}}}\left(\hat{b}_{m}^{\dagger}+\hat{b}_{m}\right)
\end{align*}
into Eq. \eqref{eq:Ham-El}, one gets the fully quantized Hamiltonian
of the $m$-th molecule, 
\begin{align}
\hat{H}_{m} & =\omega_{m}\left(\hat{b}_{m}^{\dagger}\hat{b}_{m}+\frac{1}{2}\right)\nonumber \\
 & +[\epsilon_{m}+\lambda_{m}-\omega_{m}\sqrt{s_{m}}\left(\hat{b}_{m}^{\dagger}+\hat{b}_{m}\right)]\hat{B}_{m}^{\dagger}\hat{B}_{m}.
\end{align}
Here the Huang--Rhys factor is defined as $s_{m}\equiv\lambda_{m}/\omega_{m}$.
This brings the vibrational ladder of states in the electronic ground
state 
\begin{equation}
|\mathrm{g}_{i}^{m}\rangle=\frac{\left(\hat{b}_{m}^{\dagger}\right)^{i}}{\sqrt{i!}}|0\rangle\label{eq:basis1-1}
\end{equation}
and in the electronic excited state
\begin{equation}
|\mathrm{e}_{i}^{m}\rangle\equiv\hat{B}_{m}^{\dagger}|\mathrm{g}_{i}^{m}\rangle=\hat{B}_{m}^{\dagger}\frac{\left(\hat{b}_{m}^{\dagger}\right)^{i}}{\sqrt{i!}}|0\rangle.\label{eq:basis1-1-1}
\end{equation}
$|0\rangle$ is the vacuum state in terms of electronic and vibrational
excitations.

\subsection{Hamiltonian of the vibrational aggregate}

The Hamiltonian of an aggregate of realistic molecules involves three
components: electronic states, vibrational structure for each electronic
state and the Coulomb coupling between all electronic and vibronic
levels. The first two are described by extending the Hamiltonian of
a single molecule into the space of a set of molecules within the
Heitler--London approximation, which assumes that the aggregate states
are constructed from the direct products of the molecular single excitations
\cite{Davydov-book,Amerongen2000,May2011,Valkunas2013}. We consider
only single and double excitations. The Coulomb coupling between the
$m$-th and $n$-th molecule is denoted by the resonant electronic
coupling constant $J_{mn}$ and the corresponding term is as follows:
\begin{equation}
\hat{H}_{{\rm Coulomb}}=\sum_{m\ne n}J_{mn}\hat{B}_{m}^{\dagger}\hat{B}_{n}.
\end{equation}
We neglect electrostatic interactions between vibrations in the ground
state. Within this model the Hamiltonian for the vibrational aggregate
is given by 
\begin{align}
\hat{H} & =\sum_{m}\left[\epsilon_{m}+\lambda_{m}-\omega_{m}\sqrt{s_{m}}\left(\hat{b}_{m}^{\dagger}+\hat{b}_{m}\right)\right]\hat{B}_{m}^{\dagger}\hat{B}_{m}\nonumber \\
 & +\sum_{m}\omega_{m}\left(\hat{b}_{m}^{\dagger}\hat{b}_{m}+\frac{1}{2}\right)+\sum_{m\ne n}J_{mn}\hat{B}_{m}^{\dagger}\hat{B}_{n}.\label{eq:MHD}
\end{align}
Similarly as to the electronic aggregate we get bands corresponding
to electronic states, but now the ground state $|{\rm g}\rangle$
of the aggregate is not a single quantum level, but a band of vibrational
states. Thus, there are states with all chromophores in their electronic
ground states, while vibrational excitations are arbitrary: 
\begin{equation}
|\mathrm{g}_{(i_{1}i_{2}...i_{N})}\rangle\equiv|\prod_{m}{\rm g}_{i_{m}}^{m}\rangle=\left[\prod_{m}\frac{\left(\hat{b}_{m}^{\dagger}\right)^{i_{m}}}{\sqrt{i_{m}!}}\right]|0\rangle.\label{eq:basis1}
\end{equation}
Here $i_{m}$ is a quantum number of vibrational excitation of the
$m$-th molecule. Thus $|{\rm g}_{i_{m}}^{m}\rangle$ now denotes
the electronic ground state of the $m$-th molecule being in the $i_{m}$-th
vibrational level. 

The singly-excited states are obtained by assuming that one of the
molecules is in its electronic excited state, while the others are
in their arbitrary vibrational ground states. We thus get the set
of states 
\begin{equation}
|\mathrm{e}_{n,(i_{1}i_{2}...i_{N})}\rangle\equiv|{\rm e}_{i_{n}}^{n}\negmedspace\prod_{{m\atop m\ne n}}\negmedspace{\rm g}_{i_{m}}^{m}\rangle=\hat{B}_{n}^{\dagger}\negthickspace\left[\prod_{m}\frac{\left(\hat{b}_{m}^{\dagger}\right)^{i_{m}}}{\sqrt{i_{m}!}}\right]|0\rangle.
\end{equation}
The doubly-excited states are obtained similarly, 
\begin{equation}
|\mathrm{f}_{kl,(i_{1}i_{2}...i_{N})}\rangle\equiv|{\rm e}_{i_{k}}^{k}{\rm e}_{i_{l}}^{l}\negthickspace\negthickspace\prod_{{m\atop m\ne k,l}}\negthickspace\negthickspace{\rm g}_{i_{m}}^{m}\rangle=\hat{B}_{k}^{\dagger}\hat{B}_{l}^{\dagger}\negthickspace\left[\prod_{m}\frac{\left(\hat{b}_{m}^{\dagger}\right)^{i_{m}}}{\sqrt{i_{m}!}}\right]|0\rangle.\label{eq:basis3}
\end{equation}
State ordering $k<l$ is satisfied here. A complete basis set is included
into the model since all possible combinations (multi-particle states)
of vibronic and vibrational excitations are considered, cf. single-particle
approximation, where only states $|{\rm e}_{i_{n}}^{n}\prod_{{m\atop m\ne n}}{\rm g}_{0}^{m}\rangle$
are included\cite{Philpott1971,Spano2009a}. The index notation is
further simplified by introducing the $N$-component vector $\bm{i}=(i_{1}i_{3}...i_{N})$.
Then the basis states can be written as $|\mathrm{g}_{\bm{i}}\rangle$,
$|\mathrm{e}_{n,\bm{i}}\rangle$ and $|\mathrm{f}_{kl,\bm{i}}\rangle$. 

In this setup electronic and vibrational subsystems are coupled only
through term $\left(\hat{b}_{m}^{\dagger}+\hat{b}_{m}\right)\hat{B}_{m}^{\dagger}\hat{B}_{m}$
in Hamiltonian (Eq.~\eqref{eq:MHD}). It thus induces the shifts
of electronic energies by creation or annihilation of vibrational
quantum. Otherwise, electronic and vibrational subsystems are independent.
The basis set is chosen accordingly. The other basis set is possible
by using shifted vibrational excitations in the electronic excited
states \cite{Polyutov2012,eisfeld:134103}. However, our approach
gives convenient form for various matrix elements and allows us to
easily incorporate the environment as shown below. Hamiltonian of
the ground state manifold in this basis is diagonal, 
\begin{align}
H_{\bm{i},\bm{j}}^{({\rm gg})} & =\left[\sum_{m}\omega_{m}\left(i_{m}+\frac{1}{2}\right)\right]\bm{\delta}_{\bm{ij}},\label{eq:site-ground}
\end{align}
where $\delta_{\vec{ij}}\equiv\prod_{m}\delta_{i_{m}j_{m}}$. Similarly,
the Hamiltonian of singly-excited states is given by 
\begin{align}
H_{\bm{i},\bm{j}}^{({\rm e}_{n}{\rm e}_{k})} & =\delta_{nk}\left[\epsilon_{n}+\lambda_{n}+\sum_{m}\omega_{m}\left(i_{m}+\frac{1}{2}\right)\right]\delta_{\bm{ij}}\nonumber \\
 & -\delta_{nk}\omega_{n}\sqrt{s_{n}}\langle i_{n},j_{n}\rangle\prod_{{m\atop m\ne n}}\delta_{i_{m}j_{m}}\nonumber \\
 & +(1-\delta_{nk})J_{nk}\delta_{\bm{ij}},
\end{align}
where we have defined the vibrational wavefunction overlap $\langle i_{n},j_{n}\rangle=\sqrt{i_{n}}\delta_{i_{n},j_{n}+1}+\sqrt{j_{n}}\delta_{i_{n},j_{n}-1}$.
For the double-exciton states we have 
\begin{align}
 & H_{\bm{i},\bm{j}}^{({\rm f}_{kl}{\rm f}_{k'l'})}=\nonumber \\
 & \quad\delta_{kk'}\delta_{ll'}\left[\epsilon_{k}+\epsilon_{l}+\lambda_{k}+\lambda_{l}+\sum_{m}\omega_{m}\left(i_{m}+\frac{1}{2}\right)\right]\bm{\delta}_{\bm{ij}}\nonumber \\
 & \quad-\delta_{kk'}\delta_{ll'}\omega_{k}\sqrt{s_{k}}\langle i_{k},j_{k}\rangle\prod_{{m\atop m\ne k}}\delta_{i_{m}j_{m}}\nonumber \\
 & \quad+\delta_{kk'}\delta_{ll'}\omega_{l}\sqrt{s_{l}}\langle i_{l},j_{l}\rangle\prod_{{m\atop m\ne l}}\delta{}_{i_{m}j_{m}}\nonumber \\
 & \quad+\left[\delta_{kk'}(1-\delta_{ll'})J_{ll'}+\delta_{ll'}(1-\delta_{kk'})J_{kk'}\right]\bm{\delta}_{\bm{ij}}.\label{eq:site-doubly-excited}
\end{align}

The exciton energies (eigenstate basis) $\varepsilon_{{\rm e}}$ and
$\varepsilon_{{\rm f}}$ are obtained by numerically diagonalizing
matrices defined above. The bands of singly- and doubly-excited states
are however much more complicated than those of the electronic aggregate
due to coupling between the singly-excited vibronic subbands. In the
eigenstate basis all these substates become mixed. The unitary transformation
to the eigenstate basis is thus as follows:
\begin{align}
|\mathrm{e}_{p}\rangle & =\sum_{n}\sum_{\vec i}\psi_{p,\bm{i}}^{n}|\mathrm{e}_{n,\bm{i}}\rangle,\label{eq:singly-excited}\\
|\mathrm{f}_{r}\rangle & =\sum_{{kl\atop k<l}}\sum_{\vec i}\Psi_{r,\bm{i}}^{kl}|\mathrm{f}_{kl,\bm{i}}\rangle.\label{eq:doubly-excited}
\end{align}
Note that for high vibronic numbers $i,j$ the Franck--Condon parameter
becomes small and these states do not contribute to the spectra. In
general, if one includes $\nu$ vibrational levels in description
of each of $N$ molecules, this results in $N\nu^{N}$ singly-excited
states, and $N(N-1)\nu^{N}/2$ doubly-excited states, enumerated by
indices $p$ and $r$ in the previous expressions, respectively.

For electronic excitations we consider the dipole operator defined
as 
\begin{equation}
\hat{\vec P}=\sum_{m}^{N}\bm{d}_{m}(\hat{B}_{m}^{\dagger}+\hat{B}_{m}),
\end{equation}
where $\vec d_{m}$ is the electronic transition dipole vector of
the $m$-th molecule. This form essentially reflects the Frank--Condon
approximation where the electronic transition is not coupled to vibrational
system. The dipole moments representing transitions from the ground
state to singly-excited states and from singly-excited state to the
doubly-excited states are given by
\begin{align*}
\vec{\mu}_{{\rm g}_{\bm{i}}}^{{\rm e}_{p}} & =\langle{\rm g}_{\vec i}|\hat{\vec P}|{\rm e}_{p}\rangle=\sum_{m}^{N}\vec d_{m}\psi_{p,\bm{i}}^{m}
\end{align*}
and
\begin{align*}
\vec{\mu}_{{\rm e}_{p}}^{{\rm f}_{r}} & =\langle{\rm e}_{p}|\hat{\vec P}|{\rm f}_{r}\rangle=\sum_{m,n}^{N}\sum_{\vec i}\vec d_{k}\psi_{p,\bm{i}}^{n}\Psi_{r,\bm{i}}^{(mn)}.
\end{align*}
The transition amplitudes thus have the mixed electronic--vibronic
nature encoded in eigenvectors $\psi_{p,\bm{i}}^{n}$ and $\Psi_{r,\bm{i}}^{(mn)}$.

\subsection{Coupling to the bath}

We next include the relaxation using a microscopic dephasing theory,
based on the linear coupling of the vibronic coordinate to the harmonic
overdamped bath\cite{Abramavicius2009}. Hence, we assume that the
vibronic coordinate is damped. The bath is described as a set $\left\{ \alpha\right\} $
of harmonic oscillators, whose Hamiltonian is: 
\begin{eqnarray}
\hat{H}_{{\rm B}} & = & \sum_{\alpha}\frac{1}{2}\hat{p}_{\alpha}^{2}+\frac{1}{2}w_{\alpha}^{2}\hat{x}_{\alpha}^{2}.
\end{eqnarray}
Here $\hat{p}_{\alpha}$ is the momentum and $\hat{x}_{\alpha}$ is
the coordinate operators and $w_{\alpha}$ is the frequency of the
$\alpha$-th bath oscillator. The system--bath interaction is then
given in the bilinear form 
\begin{equation}
\hat{H}_{{\rm SB}}=\sum_{m\alpha}z_{m\alpha}\hat{x}_{\alpha}\hat{q}_{m}=\sum_{m\alpha}\sqrt{\frac{z_{m\alpha}^{2}}{2\omega_{m}}}\hat{x}_{\alpha}\left(\hat{b}_{m}^{\dagger}+\hat{b}_{m}\right).\label{eq:sbc-1}
\end{equation}
We add these two operators to complete the Hamiltonian in Eq.~\eqref{eq:MHD}.
Thus, the off-diagonal fluctuations of vibronic levels translate into
diagonal fluctuations of the electronic-only aggregates due to electronic
excitation creation/annihilation operators in Eq.~\eqref{eq:sbc-1}.
More explicitly, coupling $z$ induces the vibronic off-diagonal couplings
and causes vibrational intramolecular relaxation. Resonance intermolecular
interaction $J$ will extend into the electronic energy relaxation
between different molecules. The non-zero fluctuating matrix elements
in the site basis (Eqs.~\eqref{eq:basis1}--\eqref{eq:basis3}) are
very simple: 
\begin{align}
\left(\hat{H}_{{\rm SB}}\right)_{\bm{i},\bm{j}}^{({\rm gg})} & =\langle{\rm g}_{\vec i}|\hat{H}_{{\rm SB}}|{\rm g}_{\vec j}\rangle=\mathcal{H}(\bm{i},\bm{j}),\\
\left(\hat{H}_{{\rm SB}}\right)_{\bm{i},\bm{j}}^{({\rm e}_{n}{\rm e}_{k})} & =\langle{\rm e}_{n,\vec i}|\hat{H}_{{\rm SB}}|{\rm e}_{k,\vec j}\rangle=\delta_{nk}\mathcal{H}(\bm{i},\bm{j}),\\
\left(\hat{H}_{{\rm SB}}\right)_{\bm{i},\bm{j}}^{({\rm f}_{kl}{\rm f}_{k'l'})} & =\langle{\rm f}_{kl,\vec i}|\hat{H}_{{\rm SB}}|{\rm f}_{k'l',\vec j}\rangle=\delta_{kk'}\delta_{ll'}\mathcal{H}(\bm{i},\bm{j}).
\end{align}
Here we defined an auxiliary function of bath--space fluctuations
\begin{equation}
\mathcal{H}(\bm{i},\bm{j})=\sum_{m\alpha}\sqrt{\frac{z_{m\alpha}^{2}}{2\omega_{m}}}\langle i_{m},j_{m}\rangle\hat{x}_{\alpha}\prod_{{s\atop s\ne m}}\delta_{i_{s}j_{s}}.\label{eq:aux_H}
\end{equation}
Notice, that interband fluctuations are absent, so the interband relaxation
(electronic relaxation to the ground state) is not included. Transformation
to the eigenstate basis yields the fluctuations of the eigenstate
characteristics. In the ground state manifold we have eigenstates
equivalent to the site basis since the corresponding Hamiltonian is
diagonal (Eq.~\eqref{eq:site-ground}). For the manifold of singly-excited
states we get 
\begin{equation}
\left(\hat{H}_{{\rm SB}}\right)_{p_{1}p_{2}}^{({\rm ee})}=\sum_{m}^{N}\sum_{\bm{i},\bm{j}}\psi_{p_{1},\bm{i}}^{m\ast}\psi_{p_{2},\bm{j}}^{m}\mathcal{H}(\bm{i},\bm{j}),
\end{equation}
and for the manifold of doubly-excited states 
\begin{equation}
\left(\hat{H}_{{\rm SB}}\right)_{r_{1}r_{2}}^{({\rm ff})}=\sum_{{m,n\atop m>n}}^{N}\sum_{\bm{i},\bm{j}}\Psi_{r_{1},\bm{i}}^{(mn)\ast}\Psi_{r_{2},\bm{j}}^{(m_{1}n_{1})}\mathcal{H}(\bm{i},\bm{j}).
\end{equation}

The quantities of interest, which describe the relaxation properties,
are the correlation functions of fluctuating Hamiltonian elements.
Firstly, we assume that fluctuations of different chromophores are
independent. Therefore, we can sort out and associate the bath coordinates
to specific molecules. Since the bath oscillators are independent,
correlation functions of the operator in the Heisenberg representation
with respect to the thermal equilibrium are uncorrelated, $\langle\hat{x}_{\alpha}(t)\hat{x}_{\beta}(0)\rangle=\delta_{\alpha\beta}\langle\hat{x}_{\alpha}(t)\hat{x}_{\alpha}(0)\rangle$,
and we can obtain separated baths of different molecules. Secondly,
we assume that the different molecules have statistically the same
surroundings, so the system--bath coupling is fully characterized
by the following single fluctuation correlation function: 
\begin{equation}
C_{0}(t)=\sum_{m\alpha}\frac{z_{m\alpha}^{2}}{2\omega_{m}}\langle\hat{x}_{\alpha}(t)\hat{x}_{\alpha}(0)\rangle.
\end{equation}
For the infinite number of bath oscillators they can be conveniently
expressed using the spectral density $\mathcal{C}^{\prime\prime}\left(\omega\right)$
\cite{mukbook}:\emph{ }
\begin{equation}
C_{0}\left(t\right)=\frac{1}{\pi}\intop_{-\infty}^{+\infty}\frac{1}{1-\mathrm{e}^{-\beta\omega}}\mathrm{e}^{-\imath\omega t}\mathcal{C}^{\prime\prime}(\omega)\mathrm{d}\omega,\label{eq:genCF}
\end{equation}
where $\beta=\left(k_{{\rm B}}T\right)^{-1}$ is the inverse thermal
energy. Using these functions we get the eigenstate fluctuation correlation
functions $C_{ab,cd}(t)=\langle\left(\hat{H}_{{\rm SB}}(t)\right)_{ab}\left(\hat{H}_{{\rm SB}}\right)_{cd}\rangle=C_{0}(t)h_{ab,cd}$
for different manifolds. For the electronic ground state manifold
where a single eigenstate is equivalent to the original basis state
$|\mathrm{g}_{\bm{i}}\rangle$ it yields 
\begin{equation}
C_{\bm{i}\bm{j},\bm{k}\bm{l}}^{({\rm gg})}(t)=C_{0}(t)\sum_{m}\langle i_{m},j_{m}\rangle\langle k_{m},l_{m}\rangle\prod_{{s\atop s\ne m}}\delta_{i_{s}j_{s}}\delta_{k_{s}l_{s}}.
\end{equation}
We use the shorthand vector notations $\bm{i}_{s}^{-}=(i_{1},i_{2},...,i_{s-1},i_{s}-1,i_{s+1},...,i_{N})$
and $\bm{i}_{s}^{+}=(i_{1},i_{2},...,i_{s-1},i_{s}+1,i_{s+1},...,i_{N})$,
which allow us to explicitly write: 
\begin{align}
 & C_{\bm{i}\bm{j},\bm{k}\bm{l}}^{({\rm gg})}(t)=\\
 & \quad C_{0}(t)\sum_{s}^{N}\left\{ \sqrt{i_{s}j_{s}}\delta_{\vec i\vec j_{s}^{+}}\delta_{\vec k\vec l_{s}^{+}}+\sqrt{(i_{s}+1)j_{s}}\delta_{\vec i_{s}^{+}\vec j}\delta_{\vec k\vec l_{s}^{+}}\right.\nonumber \\
 & \quad\left.+\sqrt{i_{s}(k_{s}+1)}\delta_{\vec i\vec j_{s}^{+}}\delta_{\vec k_{s}^{+}\vec l}+\sqrt{(i_{s}+1)(k_{s}+1)}\delta_{\vec i_{s}^{+}\vec j}\delta_{\vec k_{s}^{+}\vec l}\right\} .\nonumber 
\end{align}
Similarly, one can obtain the correlation functions involving the
singly- and doubly-excited states $C_{p_{1}p_{2},p_{3}p_{4}}^{({\rm ee})}(t)$,
$C_{p_{1}p_{2},r_{1}r_{2}}^{({\rm ef})}(t)$ and $C_{r_{1}r_{2},r_{3}r_{4}}^{({\rm ff})}(t)$
(see Appendix~\ref{sec:Correlation-functions-involving} for the
corresponding expressions).

\subsection{Population transfer}

As the bath induces off-diagonal fluctuations in all three bands of
states one has to consider the population transfer inside the excited
and ground manifolds (the populations of the doubly-excited states
are never created so the transport is not relevant there). The propagator
$G_{{\rm e}_{p_{2}}{\rm e}_{p_{1}}}(t_{2})$ denotes the conditional
probability of the excitation to be transferred to state $|{\rm e}_{p_{2}}\rangle\langle{\rm e}_{p_{2}}|$
from $|{\rm e}_{p_{1}}\rangle\langle{\rm e}_{p_{1}}|$ in time $t_{2}$.
Similarly, $G_{{\rm g}_{\vec j}{\rm g}_{\vec i}}(t_{2})$ is the propagator
in the electronic ground manifold. In this model the bath is considered
as the intermolecular modes which should be Markovian while intramolecular
vibrational coordinates are considered explicitly. Hence, the Redfield
theory applies for the Markovian bath. Within the secular Redfield
theory \cite{May2011}, both types of propagators satisfy the Pauli
master equation,
\begin{equation}
\frac{\partial}{\partial t}G_{ab}(t)=\sum_{{c\atop c\ne a}}k_{a\leftarrow c}G_{cb}(t)-\sum_{{c\atop c\ne a}}k_{c\leftarrow a}G_{ab}(t).
\end{equation}
Here indices $a$, $b$ and\textbf{ $c$} can be either excited state
numbers, or vectors, indicating vibrational ground states. $k_{a\leftarrow c}$
are the transfer rates in the excited (further on denoted by $k_{{\rm e}_{p_{2}}\leftarrow{\rm e}_{p_{1}}}$)
or ground state ($k_{{\rm g}_{\vec j}\leftarrow{\rm g}_{\vec i}}$).
Using the Redfield relaxation theory, one can obtain simple expressions
for the rates
\begin{align}
k_{{\rm e}_{p_{2}}\leftarrow{\rm e}_{p_{1}}} & =h_{{\rm e}_{p_{2}}{\rm e}_{p_{1}},{\rm e}_{p_{1}}{\rm e}_{p_{2}}}C^{\prime\prime}\left(\omega_{{\rm e}_{p_{1}}{\rm e}_{p_{2}}}\right)\left[\coth\left(\frac{\beta\omega_{{\rm e}_{p_{1}}{\rm e}_{p_{2}}}}{2}\right)+1\right]
\end{align}
for the excited state population transfer. For the ground state vibrational
relaxation, one has only two subsets of nonzero terms,
\begin{equation}
k_{{\rm g}_{\vec i_{s}^{-}}\leftarrow{\rm g}_{\vec i}}=i_{s}C^{\prime\prime}(\omega_{s})\left[\coth\left(\frac{\beta\omega_{s}}{2}\right)+1\right]\label{eq:gr_1}
\end{equation}
and
\begin{equation}
k_{{\rm g}_{\vec i_{s}^{+}}\leftarrow{\rm g}_{\vec i}}=\left(i_{s}+1\right)C^{\prime\prime}(-\omega_{s})\left[\coth\left(-\frac{\beta\omega_{s}}{2}\right)+1\right].\label{eq:gr_2}
\end{equation}
With these transformation expressions now it is possible to develop
the general theory describing the spectroscopic properties of vibronic
aggregates.

\section{Results}

\begin{figure}
\includegraphics{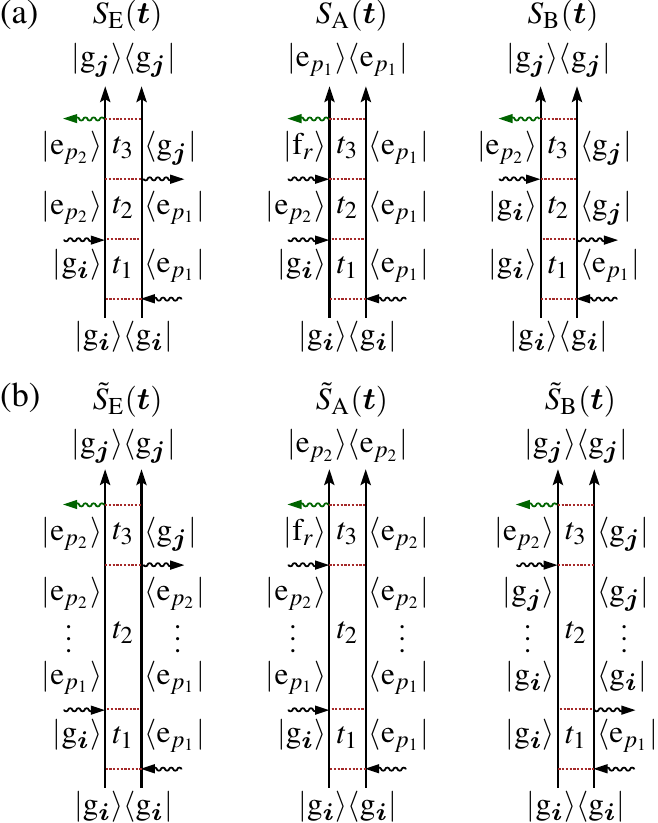}

\protect\caption{\label{fig:Double-sided-Feynman-diagrams}System response function
of the rephasing signal of the vibronic aggregate is represented by
6 double-sided Feynman diagrams without population transfer (a) and
with population transfer (b). The diagrams are denoted as responsible
for stimulated emission (E), excited state absorption (A) and ground-state
bleaching (B) processes.}
\end{figure}

In the theory of the vibronic aggregate described above we derived
all identities necessary to simulate the third-order signals in the
frame of the second-order cumulant expansion of the system response
function \cite{mukbook}. The system response function of an electronic-only
aggregate is defined as a sum of contributions (or so-called Liouville
space pathways) responsible for \emph{bleaching} of the ground state
(B), photon-induced (stimulated) \emph{emission} from the excited
state (E) and \emph{induced} \emph{absorption} of a photon in the
excited state (A). They are conveniently represented by the double-sided
Feynman diagrams, which show the system evolution during the delay
times $t_{1}$, $t_{2}$ and $t_{3}$ between the interactions. In
diagrams, ground- or excited-state populations or coherences evolve
during delay time $t_{2}$ (Fig.~\ref{fig:Double-sided-Feynman-diagrams}a).
Additionally, since population state can be transferred during $t_{2}$
in the excited state, the so-called population transfer pathways $\tilde{S}_{{\rm E}}$
and $\tilde{S}_{{\rm A}}$ are added up (Fig.~\ref{fig:Double-sided-Feynman-diagrams}b).
In the case of vibronic aggregate, this formalism has to be extended
to take into account multi-level ground state. Therefore, additional
diagrams with coherences and population transfer in the ground-state
manifold have to be included. This ingredient and the resulting final
expressions for the two-dimensional coherent spectra is described
in Appendix~\ref{sec:Spectroscopy}.

To discuss the outcomes of the developed system response function
theory for the molecular aggregate, we consider a molecular dimer
(MD) as the simplest molecular complex exhibiting vibronic phenomena,
as well as the exciton--vibrational interference. The vibrational
frequencies, site energies and Huang--Rhys factors of the constituent
molecules are taken to be the same and are denoted by $\omega_{0}\equiv\omega_{1}=\omega_{2}$,
$\epsilon\equiv\epsilon_{1}=\epsilon_{2}=1200\,\icm$ and $s=s_{1}=s_{2}$,
respectively. Also, we analyze the models in the case of weak system--bath
coupling (with Huang--Rhys factor equal to $s=0.05$) and strong coupling
($s=0.5$). Four distinct parameter sets are used and we denote the
corresponding models as \textbf{D1-D4}, indicated by stars in Fig.~\eqref{fig:(a)-Experimentally-and}.

In the \textbf{D1} model the resonant coupling constant is taken to
be $J=100$~$\icm$ and the vibrational frequency is chosen to be
$\omega_{0}=1400$~$\icm$. Such parameters are typical for the photosynthetic
pigment--protein complexes, for example, the photosynthetic antenna
of cryptophyte protein phycoerythrin 545 (the Huang--Rhys factor is
0.1) \cite{Kolli2012}. We denote this model as the weakly-coupled
P-P complex with high-frequency vibration.

In the \textbf{D2} model resonant coupling of $J=600\,\icm$ and vibrational
frequency $\omega_{0}=250\,\icm$ is used. These numbers are typical
parameters of J-aggregates, coupled to low-frequency intramolecular
vibrations. For example, in 2D electronic spectra of PVA/C8O3 tubular
J-aggregates, oscillations associated to the $160$~$\icm$ vibration
is observed and the strongest coupling between the molecules is in
a range of $640$--$1110$~$\icm$ as it was shown by Milota \emph{et
al.} \cite{Milota2013_JPCA_VibrJaggr}. In the same study, the experimental
Fourier maps were obtained. In J-aggregates the coupling to vibrations
for individual chromophores is known to decrease due to exciton delocalization
\cite{Spano2009a}. It means that, if the aggregate is approximated
as a dimer, the Huang--Rhys factor of the monomer should be multiplied
by factor of $N/2$ where $N$ is a number of chromophores in the
aggregate in the case of complete state delocalization. Therefore,
a very strong coupling to vibrations should be considered. In our
case, \textbf{D2} model with $s=0.5$ represents the typical J-aggregate
better. 

Parameters of the \textbf{D3} model ($J=100\,\icm$, $\omega_{0}=250\,\icm$)
are, as in the \textbf{D1 }model, typical for the P-P complexes. In
such molecular systems strong coupling to discrete low-frequency vibrations
are present. For example, in the measurements of two-color photon
echo of the light-harvesting complex phycocyanin-645 from cryptophyte
marine algae, long-lived oscillations possibly associated to the $194\,\icm$
vibrational mode were observed \cite{Richards2012}. Similar parameters
were also considered to be relevant for the Fenna--Mathews--Olsen
(FMO) photosynthetic light-harvesting complex \cite{Chenu2013}. Therefore,
we assume that the \textbf{D3} model effectively represents the weakly-coupled
P-P complex coupled to a low-frequency vibrational mode.

Presence of strong resonance electronic interaction between molecules
and strong coupling with high-frequency vibrations is typical for
many dimeric dyes. Hence, in the \textbf{D4 }model, the main parameters
are set to $J=600\,\icm$ and $\omega_{0}=1400\,\icm$ to be similar
to ones of perylene bisimide dye with the Huang--Rhys factor of $0.6$
\cite{Seibt2006}. 

The bath, whose degrees of freedom are not treated explicitly, is
represented by the Debye spectral density $C^{\prime\prime}(\omega)=2\lambda\omega/(\omega^{2}+\gamma^{2})$
which represents the low-frequency fluctuations. In order to get the
similar homogeneous broadening in all cases of Huang--Rhys factors,
the value of $\lambda s=25\,\icm$ is kept constant throughout all
simulations. The solvent damping energy is set to $\gamma=50$~$\icm$.
The molecular transition dipole vectors are taken to have unitary
lengths and their orientations are spread by an angle $\alpha=2\pi/5$.
Temperature is set to $150$~K ($\beta^{-1}\approx104\,\icm$).

\begin{figure}
\includegraphics{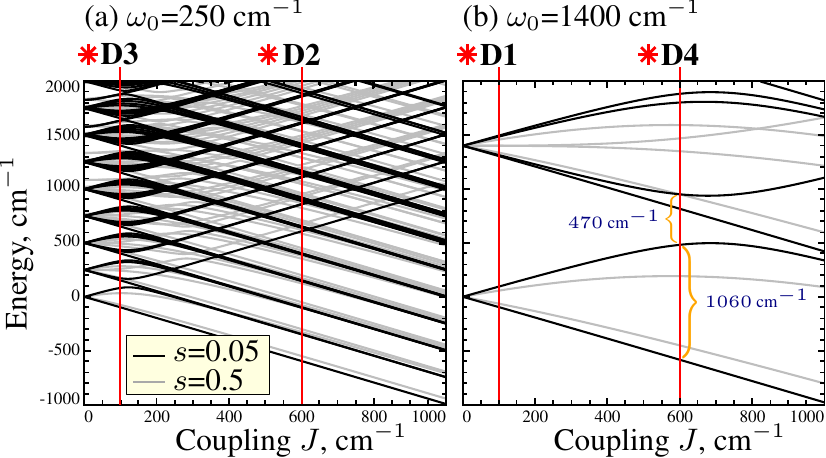}

\protect\caption{\label{fig:Dependencies-of-the}Dependencies of the singly-excited
state energies on the electronic resonance interaction ($J$ coupling)
in the case of the vibrational frequency $\omega_{0}=250\,\icm$ (a)
and $1400\,\icm$ (b). The Huang--Rhys factor is $s=0.05$ (black
lines) and $s=0.5$ (gray lines). Resonant coupling constants corresponding
to models \textbf{D1-D4} are indicated by the red vertical lines.}
\end{figure}

Let us consider the manifold of singly-excited states of all \textbf{D1}--\textbf{D4}
models. It consists of superpositions of electronic singly-excited
states and vibrational excitations of the constituent molecules. The
energy dependence on the resonant coupling constant reveals a complex
composition of the states within the singly-excited state manifold
(Fig.~\ref{fig:Dependencies-of-the}). For uncoupled molecules ($J=0$)
the ladder-type pattern of vibrational energy states is present as
the energies are equally separated by $\omega_{0}$. Increasing coupling
produces the excitonic splitting which can be seen as the red shift
of the lowest energy state and appearance of two ladder-type progressions.
However, the interaction of vibronic and electronic states induces
repulsion of the energy levels, which is mostly evident where the
ladders experience crossing, i.e. in the vicinity of the so-called
avoided crossing regions \cite{Polyutov2012}. We denote the corresponding
parameters for which the crossings occur as the exciton--vibronic
resonances. The complete mixing of the electronic and vibronic substates
is obtained for these resonances. The energy level repulsion effect
is more pronounced in the case of $s=0.5$ (see the gray lines in
Fig.~\ref{fig:Dependencies-of-the}).

In models \textbf{D1 }and \textbf{D2 }the vibrational frequency $\omega_{0}$
and resonant coupling constant $J$ differs significantly and we are
reasonably away from the resonance as can be seen in Figures~\ref{fig:(a)-Experimentally-and}
and \ref{fig:Dependencies-of-the} (the corresponding resonant coupling
values are indicated by vertical lines in the later one). Therefore,
these models can be considered as rather pure systems of vibrational
and electronic aggregates, respectively. On the contrary, parameters
of the \textbf{D3 }and \textbf{D4} models assure that the system is
very close to the exciton-vibronic resonances and the spectroscopic
signals will be more complex due to mixing. 

\begin{figure}
\includegraphics{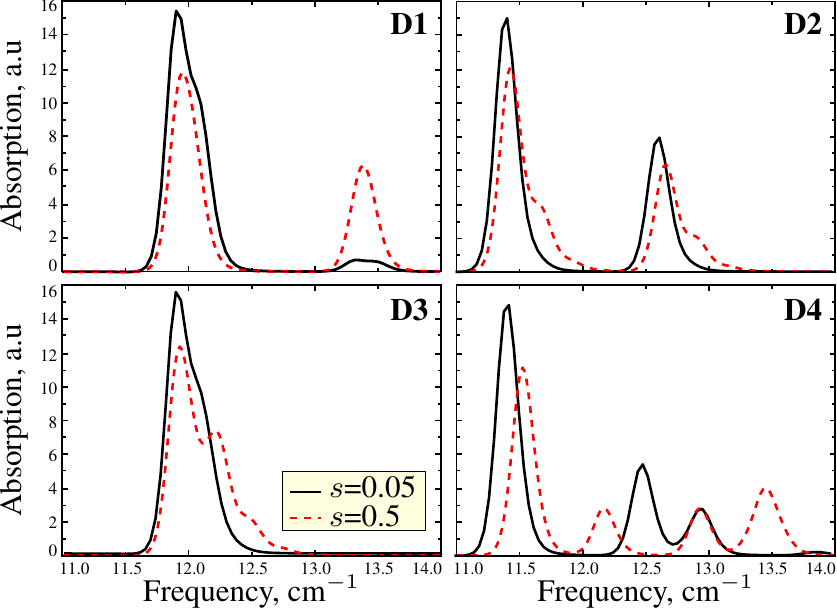}

\protect\caption{\label{fig:Absorption-spectra-of}Absorption spectra of dimers \textbf{D1-D4}
in case of Huang--Rhys factors $s=0.05$ (black solid line) and $s=0.5$
(red dashed line).}
\end{figure}

Properties of the model dimers are reflected in linear absorption
spectra (Fig.~\ref{fig:Absorption-spectra-of}). The \textbf{D1}
system has intermolecular coupling of the same order as the absorption
linewidth. Hence, both electronic transitions (and excitonic splitting)
become hidden inside the single peak 12000 cm$^{-1}$ when the Huang--Rhys
factor is $s=0.05$. Another peak at $\sim13500\,\icm$ comes from
the one-quantum level of the vibrational progression and becomes stronger
for $s=0.5$ (red dashed line). 

The \textbf{D2} model is completely opposite to the \textbf{D1}. The
excitonic splitting is large and two absorption peaks approximately
at $11500\,\icm$ and $12700\,\icm$ show the excitonic system character.
As the vibrational frequency is small, we find the vibrational progression
on both excitonic lines dependent on the Huang--Rhys factor. The \textbf{D1}
and \textbf{D2} systems, more or less, behave \char`\"{}additively\char`\"{}
where the excitonic contributions and the vibrational progressions
add up in absorption. 

Models \textbf{D3} and \textbf{D4} are very different. In the \textbf{D3},
both parameters, the excitonic resonance interaction and the vibrational
frequency, are small and the absorption spectrum shows a single broad
line at $\sim12000\,\icm$. While excitonic and vibrational contributions
are mixed, as shown in Fig. \ref{fig:Dependencies-of-the}a, surprisingly,
the absorption spectrum is relatively simple with a single electronic
peak shaped by the vibrational progression. However the shape is strongly
dependent on the Huang--Rhys factor: for $s=0.05$, one can guess
two excitonic bands (black solid line), while for $s=0.5$, the excitonic
spectrum disappears and the vibrational progression is observed. 

The fine features of mixed system is better seen in the model \textbf{D4},
which has large energy splittings between levels compared to the \textbf{D3}.
The \textbf{D4} model shows non-trivial spectrum even for small value
of the HR factor. There is a single lower-excitonic peak at 11500
cm$^{-1}$, but the higher-excitonic peak is split into two ($\sim12500\,\icm$
and $\sim13000\,\icm$). The large HR factor makes the spectrum even
more complicated where we find four peaks and they all are due to
superpositions of vibrational and electronic nature. Hence both \textbf{D3}
and \textbf{D4} systems reflect the mixed \emph{vibronic} features
of the molecular dimer. 

\begin{figure}
\includegraphics{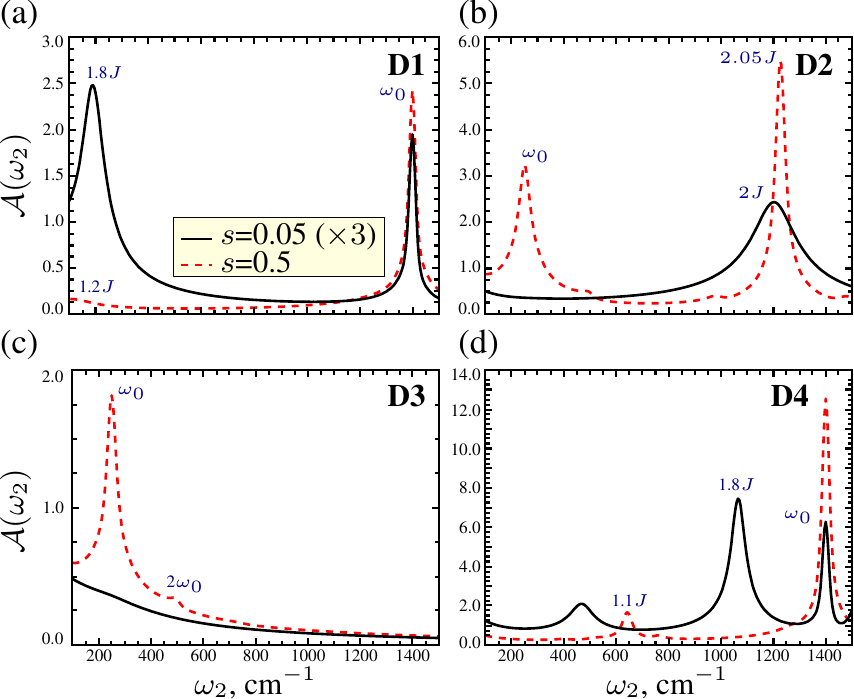}

\protect\caption{\label{fig:Maximum-amplitudes-of}Maximum of the Fourier amplitudes
characterizing oscillations in the 2D spectra of the model dimers
\textbf{D1}--\textbf{D4} (a--d panels, respectively) in case of Huang--Rhys
factors $s=0.05$ (black solid lines) and $s=0.5$ (red dashed lines).
The frequency values of the peaks are indicated in the graph.}
\end{figure}

The two-dimensional electronic spectroscopy has been suggested as
being able to distinguish between the origin of transitions of such
systems. So we analyze transition types,which could be resolved by
means of this spectroscopy for our models \textbf{D1}--\textbf{D4}.
The 2D spectra reveal as a set of peaks -- all of them contain oscillatory
contributions in the population delay time. The so-called Fourier
maps are useful for the analysis of the origin of the oscillations
in 2D spectra \cite{Butkus-Zigmantas-Abramavicius-Valkunas-CPL2012,Milota2013_JPCA_VibrJaggr,Christensson2013,Seibt2013,Calhoun2009,Panitchayangkoon2011,Turner2012}.
Thus, the maps are calculated by fitting the evolution of each point
of the 2D spectrum by the exponentially decaying function and performing
the Fourier transform of the residuals over the delay interval $t_{2}$,
\begin{equation}
A(\left|\omega_{1}\right|,\omega_{2},\omega_{3})=\intop_{0}^{\infty}\emath^{-\imath\omega_{2}t_{2}}S_{{\rm residuals}}(\left|\omega_{1}\right|,t_{2},\omega_{3})\d t_{2}.
\end{equation}
The amplitude and phase which completely describe the oscillations
of every point of the 2D spectrum are then extracted from the complex
function $A(\left|\omega_{1}\right|,\omega_{2},\omega_{3})$. As the
dependence of the amplitude on frequency $\omega_{2}$ oscillation
is available for every point of $\omega_{1}$ and $\omega_{3}$, we
suggest first to introduce a representative variable that would characterize
which oscillation frequencies are important, in general. The maximum
of the Fourier amplitude as a function of $\omega_{2}$ can be used
for that:
\begin{align*}
\mathcal{A}(\omega_{2}) & =\mbox{\ensuremath{\max}}\left[{\rm Abs}\, A(\left|\omega_{1}\right|,\omega_{2},\omega_{3})\right]_{\omega_{2}={\rm const.}}.
\end{align*}
The $\mathcal{A}(\omega_{2})$ dependencies on the oscillation frequency
for the \textbf{D1-D4} models are depicted in Fig.~\ref{fig:Maximum-amplitudes-of}
and the Fourier maps of several dominant frequencies are presented
in Figures~\ref{fig:osci-d1_d2} and \ref{fig:osci-d3_d4}. We next
discuss the models separately.

\subsection{D1 model. Weakly-coupled P-P complex with high-frequency vibration}

\begin{figure*}
\includegraphics{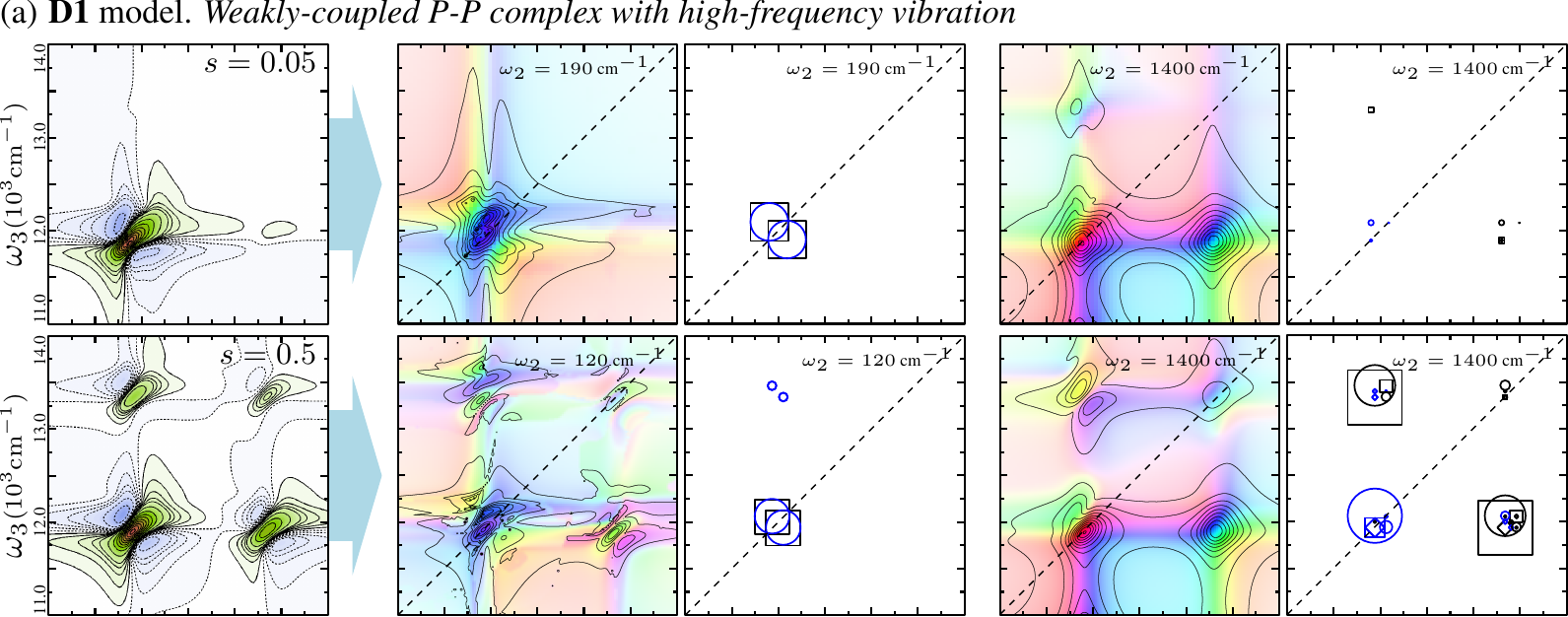}

\medskip{}

\includegraphics{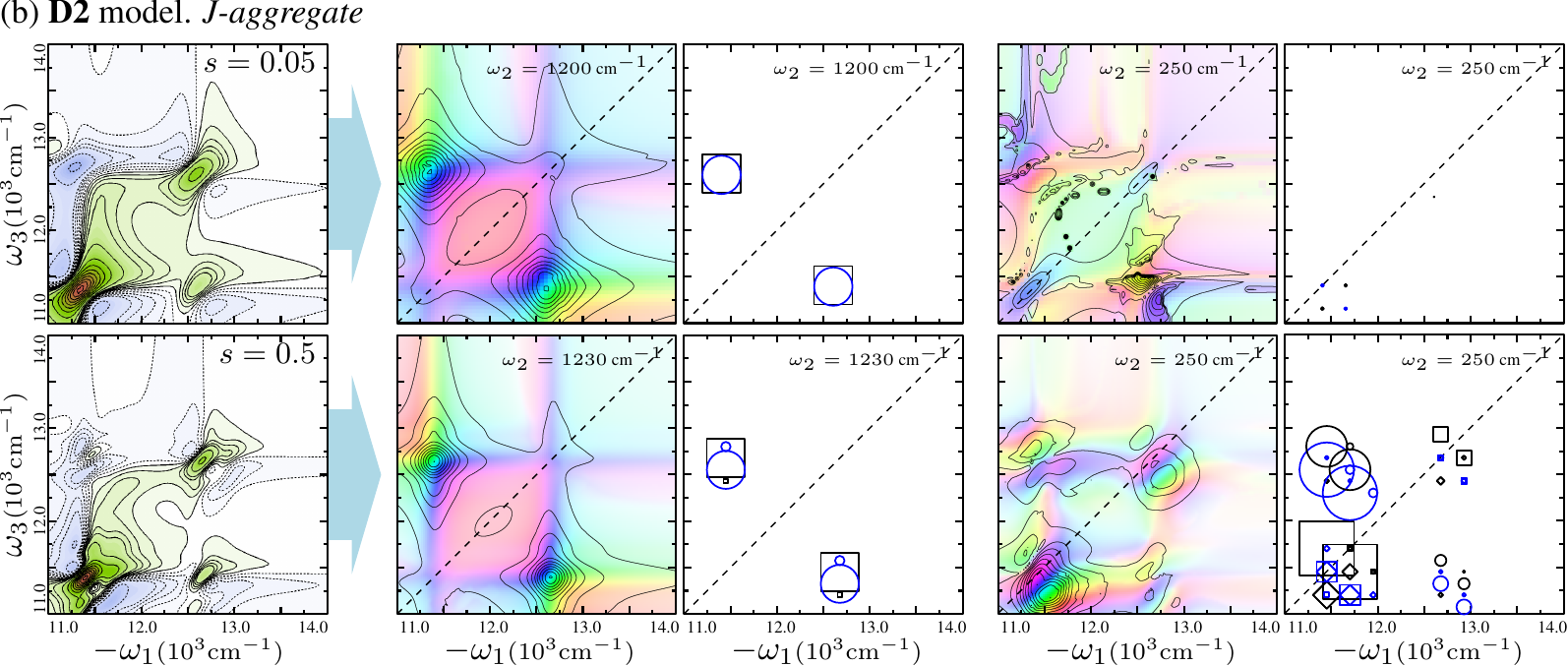}

\protect\caption{\label{fig:osci-d1_d2}Oscillations in 2D spectra of weakly-coupled
P-P complex with high-frequency vibration (\textbf{D1} model) and
J-aggregate (\textbf{D2 }model) in case of weak and strong coupling
to vibrations ($s=0.05$ and $s=0.5$, respectively). 2D rephasing
spectra at $t_{2}=0$ and two most significant Fourier maps are represented
in rows of every model with. Schemes of the oscillations-providing
contributions ($\circ$ -- excited state absorption, $\square$ --
stimulated emission and $\diamond$ -- ground state bleaching) are
presented next to the maps. The size of the symbols are proportional
to the amplitude of the corresponding contribution.}
\end{figure*}

Two dominant frequencies of $190\,\icm$ and $1400\,\icm$ representing
oscillations in spectra of the \textbf{D1} system are resolved when
$s=0.05$ (Fig.~\ref{fig:Maximum-amplitudes-of}a). Hence we consider
Fourier maps at these two frequencies. The former corresponds to the
excitonic energy splitting, but the frequency is smaller than $2J$
($190\,\icm\approx1.8J$) due to slight energy level repulsion, present
even away from the exciton--vibronic resonance. The map for $\omega_{2}=190\,\icm$
is typical for electronic coherence, as the oscillating behavior corresponding
to the excited state absorption and ground state bleaching contributions
are positioned symmetrically with respect to the diagonal line and
the oscillations are in-phase (Fig.~\ref{fig:osci-d1_d2}a). Since
the distance between the positions is smaller than the homogeneous
linewidth, the most intensive oscillations are present on the diagonal
due to constructive interference. The Fourier map at $\omega_{2}=1400\,\icm$
is a typical reflection of the vibrational/vibronic coherence as the
oscillations are present both on the diagonal line and on the cross-peaks,
characterized by complex phase dependence \cite{Butkus2013}. The
phase of oscillations is shifted by $\pi$ at the center of the lower
diagonal peak compared to the centers of the other peaks, which is
also typical for beatings of vibrational/vibronic coherences \cite{Butkus-Zigmantas-Abramavicius-Valkunas-CPL2012}.
Two more off-diagonal oscillating features at around $\omega_{3}=10500\,\icm$
are out of bounds in presented Fourier maps, hence they would be off-resonant
in a typical experiment. 

Increasing the Huang--Rhys factor to $s=0.5$ causes stronger mixing
in the system. The shape of the Fourier map at $\omega_{2}=\omega_{0}$
does not change notably, however, its amplitude increases by factor
of $~3$. The Fourier map at $\omega_{2}=120\,\icm\approx1.2J$ closely
resembles the map at $\omega_{2}=190\,\icm$ when $s=0.05$. Additional
contributions of the excited state absorption appear above the diagonal.
Features in this map are not very smooth since the lifetime of oscillations
is much shorter (note that symbol sizes in Fig.~\ref{fig:Maximum-amplitudes-of}a,
representing the amplitudes of contributions in the schemes for $s=0.05$
and $s=0.5$ are, however, similar).

\subsection{D2 model. J-aggregate.}

The strongest frequencies for model \textbf{D2} are 250~$\icm$ and
1250~$\icm$ (Fig. \ref{fig:Maximum-amplitudes-of}b). The Fourier
map of the \textbf{D2 }system at $\omega_{2}=\omega_{0}=250\,\icm$
clearly shows the large contribution from the ground state and excited
state vibrations on the diagonal and less significant excited state
absorption features in the cross-peaks (Fig.~\ref{fig:osci-d1_d2}b).
The oscillations are more intensive below the diagonal, which is consistent
with the maps of the above-mentioned PVA/C8O3 J-aggregate\cite{Milota2013_JPCA_VibrJaggr}.
If compared, the relative intensities of oscillations associated with
electronic ($\omega_{2}\approx2J$) and vibrational ($\omega_{2}\approx\omega_{0}$)
transition, one would find that the relative intensity of electronic
coherences has a tendency to decrease when increasing the Huang--Rhys
factor. Thus, for $s\gg1$, maps would be completely dominated by
the vibrational coherences. The maps at $\omega_{2}=1200\,\icm$ and
$\omega_{2}=1230\,\icm$ in Fig.~\ref{fig:osci-d1_d2}b are typical
for electronic coherences as the oscillations are diagonal-symmetric.
Note that the energy splitting is much larger than the homogeneous
linewidth and the two peaks in the maps are distinguished, cf. to
the corresponding Fourier maps in the \textbf{D1} model.

\subsection{D3 model. Weakly-coupled P-P complex with low-frequency vibration}

\begin{figure*}
\includegraphics{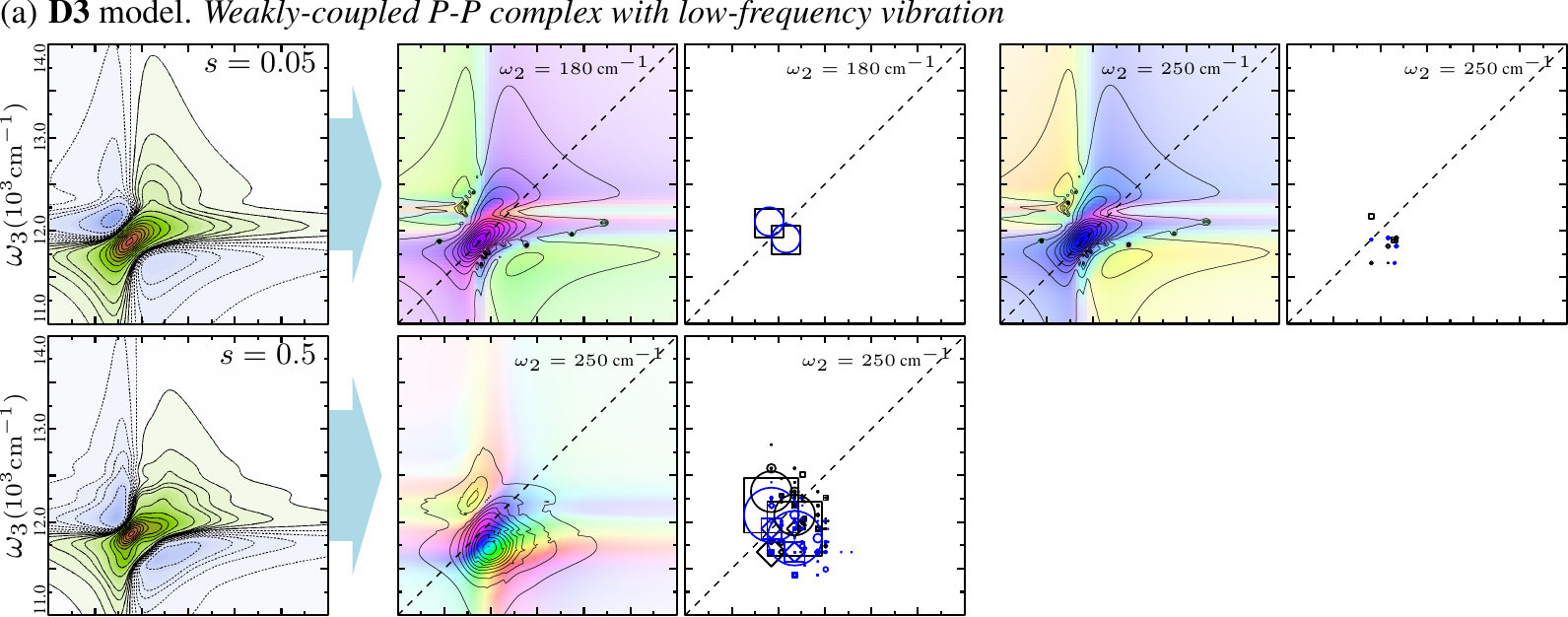}

\medskip{}

\includegraphics{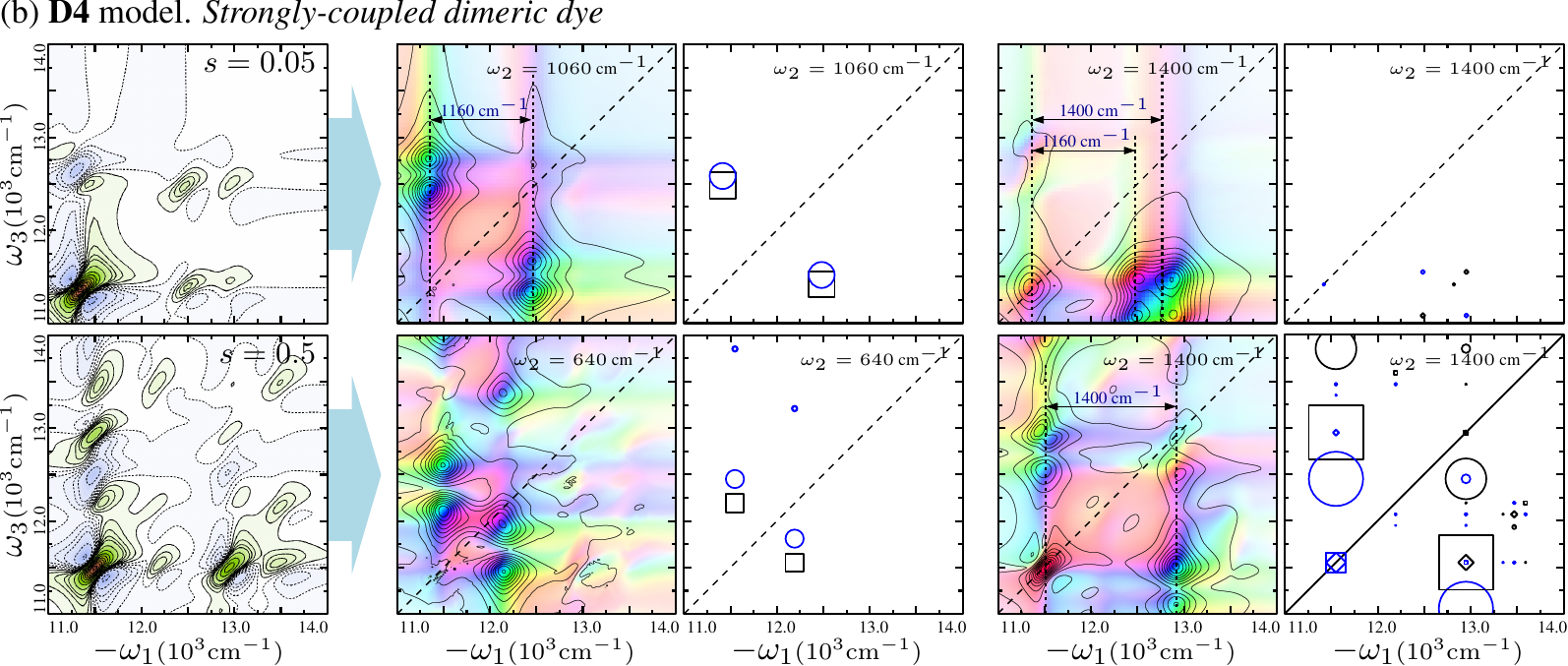}

\protect\caption{\label{fig:osci-d3_d4}Oscillations of a weakly-coupled P-P complex
with low-frequency vibration (\textbf{D3} model) and strongly-coupled
dimeric dye (\textbf{D4 }model). Presentation is analogous to Fig.~\ref{fig:osci-d1_d2}.}
\end{figure*}

Assignment of oscillations in the \textbf{D3 }with strongest peaks
shown in Fig. \ref{fig:Maximum-amplitudes-of}c is complicated since
the parameters are close to the exciton-vibronic resonance (Fig.~\ref{fig:Dependencies-of-the}a).
It might appear that there is only a continuum of low-frequency oscillations
in the spectra for $s=0.05$ since the maximum amplitude dependence
on the frequency does not contain any peaks (Fig.~\ref{fig:Maximum-amplitudes-of}c).
However, there are short-lived oscillations at $\omega_{2}=180\,\icm$
and $\omega_{2}=250\,\icm$, but their Fourier maps are not distinguishable
(Fig.~\ref{fig:osci-d3_d4}a). Increasing the Huang--Rhys factor
to $s=0.5$ enhances the $\omega_{2}=\omega_{0}$ oscillation which,
as it can be seen in the scheme next to the map in Fig.~\ref{fig:osci-d3_d4}a,
is a mixture of many different contributions.

\subsection{D4 model. Strongly-coupled dimeric dye}

The absorption spectrum of the \textbf{D4 }model\textbf{ }changes
drastically when increasing the Huang--Rhys factor (Fig.~\ref{fig:Absorption-spectra-of}).
Both transition frequencies and intensities are redistributed due
to sensitivity of the energy spectrum at the avoided crossing region.
In the 2D spectra for $s=0.05$, there are 3 clearly separable long-lived
oscillations of frequencies $\omega_{2}=0.8J\approx470\,\icm$, $\omega_{2}=1.1J\approx1060\,\icm$
and $\omega_{2}=\omega_{0}$ (Fig.~\ref{fig:Maximum-amplitudes-of}d).
The later two correspond to the excitonic energy splitting and vibrational
coherence, respectively, while the $470\,\icm$ oscillation signifies
beatings between the lower and upper states in the avoided crossing
region (the corresponding energy level gaps are indicated in Fig.~\ref{fig:Dependencies-of-the}b).
For $s=0.5$ the level repulsion effect is even more pronounced, as
the gap between the lowest energy states decreases from $1.8J$ to
$1.1J$ and the gap of the avoided crossing region increases from
$0.8J$ to $1.3J$.

The Fourier maps allow us to separate electronic and vibrational coherences
in this particular mixed case. When $s=0.05$ (Fig.~\ref{fig:osci-d3_d4}b)
the Fourier map for $\omega_{2}=1060\,\icm$ is typical for electronic
coherence. The only signature of coupling to vibrations is the oscillatory
contribution of the excited state absorption appearing above the stimulated
emission. It indicates that doubly-excited state manifold is effectively
shifted up due to vibronic coupling and, therefore, the peaks are
elongated along $\omega_{3}$ axis in the Fourier maps. The $\omega_{2}=1400\,\icm$
map is exceptionally created by the ground state vibrations, however,
the energy level structure in the excited state manifold is reflected
as the distance between some oscillating features in the Fourier maps
are found to be equal to $1060\,\icm$ (see the labels with arrows
in Fig.~\ref{fig:osci-d3_d4}b).

The $\omega_{2}=1400\,\icm$ oscillation becomes mixed if $s=0.5$.
As it can be seen in the corresponding scheme of oscillations, contributions
from all types of diagrams appear and heavily congest the Fourier
map. The lowest diagonal peak becomes oscillating due to stimulated
emission and ground state bleaching contributions. The map for the
$\omega_{2}=640\,\icm$ oscillation is similar to one for $\omega_{2}=1060\,\icm$
presented above. Stronger coupling to vibrations induce appearance
of additional oscillations in the excited state manifold, seen as
two peaks above the diagonal.

\section{Discussion}

2D electronic spectroscopy is the ultimate tool capable to directly
reflect coherent system dynamics, manifested by spectral oscillations
and beatings. Frequency and decay rate of oscillations indicate the
energy difference of states involved in the coherent superposition
and the coherent state lifetime, respectively. Positions of emerging
oscillations in spectra are conveniently depicted by the use of Fourier
maps, thus providing us one more additional dimension. For example,
peaks in the maps, symmetric with respect to the diagonal, reflect
the electronic coherence evolution in the excited state (\textbf{D1
}and \textbf{D2} models in Fig.~\ref{fig:osci-d1_d2}). The information
about the phase of oscillations provides additional information about
the ongoing processes. Therefore, fitting the experimental data by
means of the simulated Fourier maps would lead to unambiguous conclusions.

\subsection{Nature of coherences}

Let us discuss about the quantum coherences in molecular aggregates.
Quantum mechanical description of \emph{electronic} excitation treats
the rigid constituent molecule's skeleton as the potential energy
surface for electrons. The resulting electron density dynamics after
photo-excitation can be approximated by the oscillatory electric dipole
moment. Thus, the coupling between the molecules produces the discrete
energy levels in the single excited manifold of the \emph{electronic}
aggregate. In a similar way, if intramolecular vibrations are considered
in an isolated molecule, the harmonic/anharmonic oscillator model
for the electronic ground and excited states can be applied. This
also results in discrete spectrum of \emph{vibrational} levels. These
two pictures merge due to the intermolecular coupling and the electronic
and vibrational subsystems mix up. It is then natural to try to quantify,
how much of electronic or vibrational character is inherited in the
composite system. However, this often leads to many ambiguities, for
example, in linear spectra of J-aggregates\cite{FidderCP2007}. 

There have been many attempts to unambiguously distinguish between
vibrational, vibronic and excitonic coherences visible as oscillations
in 2D spectra. However, the question of how to do that is proper only
if mixing in the system is low. As it was shown here, two conditions
for low mixing can be distinguished: (i) the coupling between vibrational
and electronic subsystem has to be weak (small Huang--Rhys factor);
(ii) the system must not be in a vicinity of exciton--vibronic resonance,
represented by the avoided crossing region in the energy spectrum.
These conditions are best met for high-frequency intramolecular vibrations
in weakly-coupled P--P complexes and low-frequency vibrations in strongly-coupled
aggregates, the \textbf{D1} and \textbf{D2} models, respectively.

In the case of substantial mixing of the coherences of electronic
and vibrational character, the information about the transition composition
can be evaluated from coherent oscillations in some cases. It is most
obvious in the \textbf{D2} model representing the case of the J-aggregate,
when the electronic system is strongly coupled to low-frequency vibration
(second row in Fig.~\ref{fig:osci-d1_d2}b). Despite the strong mixing
the Fourier map for the electronic frequency ($\omega_{2}=1230\,\icm$)
contains diagonal-symmetric features, similar to those present in
the weak mixing case. In the \textbf{D4 }model, which stands as an
equivalent of the strongly-coupled dimeric dye, the mixture of coherences
for $\omega_{2}=1060\,\icm$ and $\omega_{2}=1400\,\icm$ can also
be disentangled ($s=0.05$, Fig.~\ref{fig:osci-d3_d4}b). Firstly,
the Fourier map at $\omega_{2}=1060\,\icm$ contains diagonal-symmetric
peaks, which would suggest, that this particular coherence is rather
electronic. Secondly, there are features in the map at $\omega_{2}=1400\,\icm$
separated by $1400\,\icm$ and $1060\,\icm$ as well as the peak on
the diagonal exhibiting high-frequency oscillations. The later fact
as well as the obviously stronger oscillations below the diagonal
shows that the origin of the $1400\,\icm$ oscillation is rather vibrational.
The similar analysis can be applied to the \textbf{D4} model with
$s=0.5$, where the mere evidence of vibrational content is the diagonal
oscillating peak in the $\omega_{2}=1400\,\icm$ map.

One cannot discriminate between coherences of dominating electronic
or vibrational character in weakly-interacting photosynthetic complexes,
coupled to low-frequency vibrations. This is clearly demonstrated
by the \textbf{D3} model in both cases of weak and strong electron--phonon
coupling (first and second rows in Fig.~\ref{fig:osci-d3_d4}, respectively).
The coherences in the system are highly mixed and no typical patterns,
which were present in the Fourier maps of the other systems, are found
here. The Fourier map in case of strong coupling to vibrations is
composed of many contributions, evolving in the ground and excited
states (the second row in Fig.~\ref{fig:osci-d3_d4}), thus, indicating
complete state character mixing. Hence, the \emph{electronic} or \emph{vibrational}
transitions are properly qualified, while \emph{vibronic} and \emph{how-much-vibronic}
is a vague concept and should be avoided. Instead one should treat
such coherences as simply mixed, which is completely a proper concept
in quantum mechanics.

\subsection{Lifetime of coherences in aggregates }

The fact that some coherences are less visible in the Fourier maps
is to high degree related to their lifetimes. Obviously, oscillations
which decay fast will be vaguely captured by the Fourier transform
or even will not be present in the maps at all. Let us now concentrate
on the maximum of the Fourier amplitude dependence on frequency, $\mathcal{A}(\omega_{2})$,
in case of $s=0.05$, presented by the solid lines in Fig.~\ref{fig:Maximum-amplitudes-of}.
The widths of the peaks are given by the lifetime of the corresponding
oscillations. 

The lifetime of the vibrational ground state coherence depends only
on the overlap of vibrational frequency and bath spectral density.
It can be deduced from Eqs.~\eqref{eq:gr_1} and \eqref{eq:gr_2}.
For example, the lifetime of $|{\rm g}_{0}{\rm g}_{0}\rangle\langle{\rm g}_{0}{\rm g}_{1}|$
coherence $\tau_{01}=2(\gamma_{{\rm g}_{0}{\rm g}_{0}}+\gamma_{{\rm g}_{0}{\rm g}_{1}})^{-1}$
is $\sim3$~ps for $\omega_{0}=1400\,\icm$ and the width of the
corresponding peak in the $\mathcal{A}(\omega_{2})$ dependence is
$\sim40\,\icm$ (Fig.~\ref{fig:Maximum-amplitudes-of}a and d). The
lifetime of $\omega_{0}=250\,\icm$ coherence is $\sim500$~fs, thus,
the corresponding peak in $\mathcal{A}(\omega_{2})$ is very broad
and, therefore, hardly distinguishable (Fig.~\ref{fig:Maximum-amplitudes-of}b
and c). This effect essentially depends on the spectral density function
(including the shape and the amplitude) and its value at the corresponding
vibrational frequency \cite{Jankowiak_JPCB2013_SpectralDensities}.

The lifetimes of coherences in the excited state manifold are not
that trivial. On one hand, transfer rates relating electronic states
of purely electronic aggregates depend on the absolute value of the
bath spectral density at the corresponding frequency. Additionally,
they depend on extend of delocalization of a particular state. On
the other hand, transfer rates between vibronic states of a single
molecule are the same as of the ground state vibrational states. In
our case, these two pictures are merged and the lifetimes of mixed
coherences cannot be expressed in simple terms. 

It has been shown, that the lifetimes of vibronic coherences increase
significantly, if the electronic transitions are close to vibrational
frequencies even if the Huang--Rhys factor is small ($s<0.1$) \cite{Chenu2013,Christensson_JPCB2012}.
This is evident in the $\mathcal{A}(\omega_{2})$ dependencies, as
well: the lifetime of the $\omega_{2}=1.8J$ oscillation in the \textbf{D1}
model is smaller than the corresponding lifetime of the frequency
oscillation in the \textbf{D4 }model by factor of $\sim1.8$ while
the lifetimes of the $\omega_{2}=\omega_{0}$ coherences are identical.
If compared, $\omega_{2}=2J$ oscillations in the \textbf{D2 }model
decay at least $3$ times faster than the $\omega_{2}=1.8J$ oscillations
in the \textbf{D4 }model.

If the Huang--Rhys factors are large ($s\gtrsim0.5$), low-frequency
vibrational coherences in the ground state decay faster than in the
case of weak coupling to vibrations discussed above (see dashed lines
in Fig.~\ref{fig:Maximum-amplitudes-of}). This is due to the lower
value of the reorganization energy, which is $\lambda=50\,\icm$ for
$s=0.5$ (cf. $\lambda=500\,\icm$ for $s=0.05$). Stronger interaction
with vibrations induces more mixing in the system. Therefore, we can
see long-lived coherences of $\omega_{2}=2.05J$ in the \textbf{D2}
model. We can thus conclude that the electronic coherences effectively
borrow some lifetime from the vibrational coherences due to the quantum
mechanical mixing. The mixing and borrowing of the dipole strength
in excitonic systems is a well-known phenomenon, however the lifetime
borrowing is poorly discussed.

\subsection{Energetic disorder}

Energetic disorder is yet another important parameter for the coherent
state evolution. It was shown for molecular dimer with vibrations,
that in the case of substantial energy disorder, coherences of prevailing
vibrational/vibronic character will dominate over those of rather
electronic character \cite{Butkus2013}. From the discussion above
it follows that this effect will be more significant for rather pure
systems: weakly-coupled P--P complexes with high-frequency vibrations
and J-aggregates (\textbf{D1} and \textbf{D2} models, respectively).
In both cases vibrational coherences will dominate in the Fourier
maps, while the electronic coherences will dephase fast because of
combined influence of the energetic disorder and the lifetime of the
state. In the mixed systems (\textbf{D3} and \textbf{D4}) the result
will be more complicated. The reason behind is that the mixing occurs
when resonances match, i.e. at the exciton-vibronic resonance. Disorder
will make the matching less significant for most of ensemble members
show less mixing. Hence, the electronic and vibronic character for
a disordered ensemble is better defined and should be better identified
in experiments.

\section{Conclusions}

In this paper, theory of molecular aggregate with intramolecular vibrations
for coherent spectroscopy was presented. It accounts for incoherent
and coherent effects caused by excitonic coupling and exciton--vibronic
interaction. The molecular dimer model is used for simulation purposes
of typical systems in a wide range of parameters to reflect pigment--protein
complexes, J-aggregates or dimeric molecular dyes.

Regarding the question of distinguishing the electronic, vibronic
or vibrational coherences, we conclude that the question itself is
fully defined and proper only if the character of states is pure (electronic
or vibrational). We have shown that such a separation is indeed possible
for systems, where the resonant coupling and vibrational frequency
is off-resonant, i.e. the system is away from the so-called exciton--vibronic
resonance. The analysis of oscillations in mixed systems is of qualitative
significance and allows us to to tell if the specified oscillation
is of the mixed origin. The frequencies of transitions, involved in
the mixture can be resolved.

The positions of oscillating features must be taken into careful consideration
when analyzing experimental data. There are spectral regions of mixed
systems, where oscillations exceptionally due to coherences in the
ground or excited states can be found, thus providing more information
about system composition. For example, the property that features
in the Fourier maps are asymmetric with respect to the diagonal is
the signature that the corresponding state is mixed.

Lifetime of excitonic coherences is mostly determined by the coupling
to discrete modes of intramolecular vibrations. These modes, on their
own, are coupled to the continuum of low-frequency bath fluctuations,
represented by the spectral density. Thus, the overlap of the spectral
density function and frequencies of intramolecular vibrations as well
as the form of spectral density function directly influences the lifetime
of electronic coherences.

\appendix

\section{\label{sec:Correlation-functions-involving}Correlation functions
involving singly- and doubly-excited states}

Singly-excited eigenstates are obtained by unitary transformation,
and we get the same symmetry properties as for 

\begin{align}
C_{p_{1}p_{2},p_{3}p_{4}}^{({\rm ee})}(t) & =C_{0}(t)\sum_{m,n}^{N}\sum_{\bm{i},\bm{j}}\sum_{\bm{k},\bm{l}}\sum_{a}^{N}\psi_{p_{1},\bm{i}}^{m\ast}\psi_{p_{2},\bm{j}}^{m}\psi_{p_{3},\bm{k}}^{n\ast}\psi_{p_{4},\bm{l}}^{n}\nonumber \\
 & \times\langle i_{a},j_{a}\rangle\langle k_{a},l_{a}\rangle\prod_{{s\atop s\ne a}}\delta_{i_{s}j_{s}}\delta_{k_{s}l_{s}}.
\end{align}
Here the first sum is over different chromophores, the second and
third sum is over the vibrational levels of the whole aggregate and
finally the sum over $a$ is over the different vibrational modes
(which is identical to the number of sites since each site has one
vibrational coordinate). We then get the following result:\begin{widetext}
\begin{align*}
C_{p_{1}p_{2},p_{3}p_{4}}^{({\rm ee})}(t) & =C_{0}(t)\sum_{\bm{i},\bm{k}}\sum_{s}^{N}\left\{ \sqrt{i_{s}k_{s}}\xi_{\bm{i}_{s}^{-}}^{(p_{1}p_{2})}\xi_{\bm{k}_{s}^{-}}^{(p_{3}p_{4})}+\sqrt{(i_{s}+1)k_{s}}\xi_{\bm{i}_{s}^{+}}^{(p_{1}p_{2})}\xi_{\bm{k}_{s}^{-}}^{(p_{3}p_{4})}\right.\\
 & \left.+\sqrt{i_{s}(k_{s}+1)}\xi_{\bm{i}_{s}^{-}}^{(p_{1}p_{2})}\xi_{\bm{k}_{s}^{+}}^{(p_{3}p_{4})}+\sqrt{(i_{s}+1)(k_{s}+1)}\xi_{\bm{i}_{s}^{+}}^{(p_{1}p_{2})}\xi_{\bm{k}_{s}^{+}}^{(p_{3}p_{4})}\right\} .
\end{align*}
\end{widetext} Here 
\begin{equation}
\xi_{\bm{i}_{s}^{\pm}}^{(p_{a}p_{b})}=\sum_{n}^{N}\psi_{p_{a},\bm{i}}^{n\ast}\psi_{p_{b},\bm{i}_{s}^{\pm}}^{n}.
\end{equation}
 For the functions involving the double excitations we can write similarly\begin{widetext}
\begin{align*}
C_{p_{1}p_{2},r_{1}r_{2}}^{({\rm ef})}(t) & =C_{0}(t)\sum_{\bm{i},\bm{k}}\sum_{s}^{N}\left\{ \sqrt{i_{s}k_{s}}\xi_{\bm{i}_{s}^{-}}^{(p_{1}p_{2})}\Xi_{\bm{k}_{s}^{-}}^{(r_{1}r_{2})}+\sqrt{(i_{s}+1)k_{s}}\xi_{\bm{i}_{s}^{+}}^{(p_{1}p_{2})}\Xi_{\bm{k}_{s}^{-}}^{(r_{1}r_{2})}\right.\\
 & \left.+\sqrt{i_{s}(k_{s}+1)}\xi_{\bm{i}_{s}^{-}}^{(p_{1}p_{2})}\Xi_{\bm{k}_{s}^{+}}^{(r_{1}r_{2})}+\sqrt{(i_{s}+1)(k_{s}+1)}\xi_{\bm{i}_{s}^{+}}^{(p_{1}p_{2})}\Xi_{\bm{k}_{s}^{+}}^{(r_{1}r_{2})}\right\} 
\end{align*}
 and 
\begin{align*}
C_{r_{1}r_{2},r_{3}r_{4}}^{({\rm ff})}(t) & =C_{0}(t)\sum_{\bm{i},\bm{k}}\sum_{s}^{N}\left\{ \sqrt{i_{s}k_{s}}\Xi_{\bm{i}_{s}^{-}}^{(r_{1}r_{2})}\Xi_{\bm{k}_{s}^{-}}^{(r_{3}r_{4})}+\sqrt{(i_{s}+1)k_{s}}\Xi_{\bm{i}_{s}^{+}}^{(r_{1}r_{2})}\Xi_{\bm{k}_{s}^{-}}^{(r_{3}r_{4})}\right.\\
 & \left.+\sqrt{i_{s}(k_{s}+1)}\Xi_{\bm{i}_{s}^{-}}^{(r_{1}r_{2})}\Xi_{\bm{k}_{s}^{+}}^{(r_{3}r_{4})}+\sqrt{(i_{s}+1)(k_{s}+1)}\Xi_{\bm{i}_{s}^{+}}^{(r_{1}r_{2})}\Xi_{\bm{k}_{s}^{+}}^{(r_{3}r_{4})}\right\} ,
\end{align*}
\end{widetext} where 
\begin{equation}
\Xi_{\bm{i}_{s}^{\pm}}^{(r_{1}r_{2})}=\sum_{{m,n\atop m>n}}^{N}\Psi_{r_{1},\bm{i}}^{(mn)\ast}\Psi_{r_{2},\bm{i}_{s}^{\pm}}^{(mn)}.
\end{equation}

\section{\label{sec:Spectroscopy}Response functions for 2D photon echo spectra}

We consider the photon-echo signals of the 2D electronic spectroscopy
in the impulsive limit. In this limit, the laser pulses are assumed
as infinitely short and, therefore, the measured intensity of electric
field is proportional to the system response function, the expressions
of which are presented in this appendix following the notation used
in Fig.~\ref{fig:Double-sided-Feynman-diagrams} \cite{Abramavicius2009}.
The expressions for the response involve the spectral lineshape functions
$g_{ab,cd}(t)\equiv h_{ab,cd}g_{0}(t)$. These are given by the linear
integral transformation of the bath correlation functions, $g_{0}(t)=\int_{0}^{t}\d t'\int_{0}^{t'}\d t''\left\langle C_{0}(t'')C_{0}(0)\right\rangle $
\cite{mukbook}. The response functions of the photon-echo (rephasing)
signal when transport is ignored are then ($\vec t\equiv\left\{ t_{3},t_{2},t_{1}\right\} $)
\begin{align}
 & S_{{\rm B}}(\vec t)=\imath^{3}\theta(\bm{t})\sum_{\vec i\vec j}\sum_{p_{1}p_{2}}p_{{\rm g}_{\vec i}}(\delta_{\vec i,\vec j}G_{{\rm g}_{\vec i}{\rm g}_{\vec i}}(t_{2})+\zeta_{\vec i,\vec j})\\
 & \times\left\langle \vec{\mu}_{{\rm g}_{\vec i}}^{{\rm e}_{p_{1}}}\vec{\mu}_{{\rm e}_{p_{1}}}^{{\rm g}_{\vec j}}\vec{\mu}_{{\rm g}_{\vec j}}^{{\rm e}_{p_{2}}}\vec{\mu}_{{\rm e}_{p_{2}}}^{{\rm g}_{\vec i}}\right\rangle \emath^{{\rm i}\xi_{{\rm e}_{p_{1}}{\rm g}_{\vec i}}t_{1}-{\rm i}\xi_{{\rm g}_{\vec i}{\rm g}_{\vec j}}t_{2}-{\rm i}\xi_{{\rm e}_{p_{2}}{\rm g}_{\vec j}}t_{3}}\nonumber \\
 & \times\emath^{\phi_{{\rm e}_{p_{1}}{\rm g}_{\vec j}{\rm e}_{p_{2}}{\rm g}_{\vec i}}(0,t_{1},t_{1}+t_{2}+t_{3},t_{1}+t_{2})},\nonumber 
\end{align}
 
\begin{align}
 & S_{{\rm E}}(\vec t)=\imath^{3}\theta(\bm{t})\sum_{\vec i\vec j}\sum_{p_{1}p_{2}}p_{{\rm g}_{\vec i}}(\delta_{p_{1}p_{2}}G_{{\rm e}_{p_{1}}{\rm e}_{p_{2}}}(t_{2})+\zeta_{p_{1}p_{2}})\\
 & \times\left\langle \vec{\mu}_{{\rm g}_{\vec i}}^{{\rm e}_{p_{1}}}\vec{\mu}_{{\rm g}_{\vec i}}^{{\rm e}_{p_{2}}}\vec{\mu}_{{\rm e}_{p_{2}}}^{{\rm g}_{\vec j}}\vec{\mu}_{{\rm e}_{p_{2}}}^{{\rm g}_{\vec j}}\right\rangle \emath^{{\rm i}\xi_{{\rm e}_{p_{1}}{\rm g}_{\vec i}}t_{1}-\imath\xi_{{\rm e}_{p_{2}}{\rm e}_{p_{1}}}t_{2}-{\rm i}\xi_{{\rm e}_{p_{1}}{\rm g}_{\vec j}}t_{3}}\nonumber \\
 & \times\emath^{\phi_{{\rm e}_{p_{1}}{\rm g}_{\vec j}{\rm e}_{p_{2}}{\rm g}_{\vec i}}(0,t_{1}+t_{2},t_{1}+t_{2}+t_{3},t_{1})}\nonumber 
\end{align}
and 
\begin{align}
 & S_{{\rm A}}(\vec t)=-\imath^{3}\theta(\bm{t})\sum_{\vec i}\sum_{p_{1}p_{2}}\sum_{r}p_{{\rm g}_{\vec i}}(\delta_{p_{1}p_{2}}G_{{\rm e}_{p_{1}}{\rm e}_{p_{1}}}(t_{2})+\zeta_{p_{1}p_{2}})\\
 & \times\left\langle \vec{\mu}_{{\rm g}_{\vec i}}^{{\rm e}_{p_{1}}}\vec{\mu}_{{\rm g}_{\vec i}}^{{\rm e}_{p_{2}}}\vec{\mu}_{{\rm e}_{p_{2}}}^{{\rm f}_{r}}\vec{\mu}_{{\rm f}_{r}}^{{\rm e}_{p_{1}}}\right\rangle \emath^{\imath\xi_{{\rm e}_{p_{1}}{\rm g}_{\vec i}}t_{1}-\imath\xi_{{\rm e}_{p_{2}}{\rm e}_{p_{1}}}t_{2}-\imath\xi_{{\rm f}_{r}{\rm e}_{p_{1}}}t_{3}}\nonumber \\
 & \times\emath^{\phi_{{\rm e}_{p_{1}}{\rm f}_{r}{\rm e}_{p_{2}}{\rm g}_{\vec i}}(0,t_{1}+t_{2}+t_{3},t_{1}+t_{2},t_{1})}.\nonumber 
\end{align}
Here, the complex variable $\xi_{ab}=\omega_{ab}-\imath\frac{1}{2}(\gamma_{a}+\gamma_{b})$
is used to take into account the state dephasing due to finite lifetime,
$\gamma_{a}=\frac{1}{2}\sum_{a'\ne a}k_{a\leftarrow a'}$. $p_{{\rm g}_{\vec i}}$
is the Boltzmann probability for the system to be in the $\vec i$-th
vibrational state prior the excitation and $\theta(\vec t)$ is the
product of Heaviside functions, $\theta(t_{1})\theta(t_{2})\theta(t_{3})$.
The auxiliary function is
\begin{align}
 & \phi_{{\rm e}_{p_{1}}c{\rm e}_{p_{2}}{\rm g}_{\vec i}}(\tau_{4},\tau_{3},\tau_{2},\tau_{1})=\nonumber \\
 & \quad-g_{{\rm e}_{p_{1}}{\rm e}_{p_{1}}}(\tau_{43})-g_{cc}(\tau_{32})-g_{{\rm e}_{p2}{\rm e}_{p2}}(\tau_{21})\\
 & \quad+g_{{\rm e}_{p_{1}}c}(\tau_{32})+g_{{\rm e}_{p_{1}}c}(\tau_{43})-g_{{\rm e}_{p_{1}}c}(\tau_{42})\nonumber \\
 & \quad-g_{{\rm e}_{p_{1}}{\rm e}_{p_{2}}}(\tau_{32})+g_{{\rm e}_{p_{1}}{\rm e}_{p_{2}}}(\tau_{31})+g_{{\rm e}_{p_{1}}{\rm e}_{p_{2}}}(\tau_{42})\nonumber \\
 & \quad-g_{{\rm e}_{p_{1}}{\rm e}_{p_{2}}}(\tau_{41})+g_{c{\rm e}_{p_{2}}}(\tau_{21})+g_{c{\rm e}_{p_{2}}}(\tau_{32})-g_{c{\rm e}_{p_{2}}}(\tau_{31})\nonumber \\
 & \quad-g_{c{\rm g}_{\vec i}}(\tau_{21})+g_{c{\rm g}_{\vec i}}(\tau_{24})+g_{c{\rm g}_{\vec i}}(\tau_{31})-g_{c{\rm g}_{\vec i}}(\tau_{34}),\nonumber 
\end{align}
where $c$ stands for either doubly-excited state ${\rm f}_{r}$,
either ground state ${\rm g}_{\vec j}$. Response function components
with transport are
\begin{align}
 & \tilde{S}_{{\rm B}}(\vec t)=\imath^{3}\bm{\theta}(\bm{t})\sum_{\vec i\vec j}\sum_{p_{1}p_{2}}p_{{\rm g}_{\vec i}}\zeta_{\vec i\vec j}G_{{\rm g}_{\vec j}{\rm g}_{\vec i}}(t_{2})\\
 & \times\left\langle \vec{\mu}_{{\rm g}_{\vec i}}^{{\rm e}_{p_{1}}}\vec{\mu}_{{\rm e}_{p_{1}}}^{{\rm g}_{\vec i}}\vec{\mu}_{{\rm g}_{\vec j}}^{{\rm e}_{p_{2}}}\vec{\mu}_{{\rm e}_{p_{2}}}^{{\rm g}_{j}}\right\rangle \emath^{\imath\xi_{{\rm e}_{p_{1}}{\rm g}_{\vec i}}t_{1}-\imath\xi_{{\rm e}_{p_{2}}{\rm g}_{\vec j}}t_{3}+\varphi_{{\rm e}_{p_{2}}{\rm g}_{\vec j}{\rm e}_{p_{2}}{\rm e}_{p_{1}}}^{\ast}(\vec t)},\nonumber 
\end{align}
 
\begin{align}
 & \tilde{S}_{{\rm E}}(\vec t)=\imath^{3}\bm{\theta}(\bm{t})\sum_{\vec i\vec j}\sum_{p_{1}p_{2}}p_{{\rm g}_{\vec i}}\zeta_{p_{1}p_{2}}G_{{\rm e}_{p_{2}}{\rm e}_{p_{1}}}(t_{2})\\
 & \times\left\langle \vec{\mu}_{{\rm g}_{\vec i}}^{{\rm e}_{p_{1}}}\vec{\mu}_{{\rm g}_{\vec i}}^{{\rm e}_{p_{1}}}\vec{\mu}_{{\rm e}_{p_{2}}}^{{\rm g}_{\vec j}}\vec{\mu}_{{\rm e}_{p_{2}}}^{{\rm g}_{\vec j}}\right\rangle \emath^{\imath\xi_{{\rm e}_{p_{1}}{\rm g}_{\vec i}}t_{1}-\imath\xi_{{\rm e}_{p_{2}}{\rm g}_{\vec j}}t_{3}+\varphi_{{\rm e}_{p_{2}}{\rm g}_{\vec j}{\rm e}_{p_{2}}{\rm e}_{p_{1}}}^{\ast}(\vec t)},\nonumber 
\end{align}
and 
\begin{align}
 & \tilde{S}_{{\rm A}}(\vec t)=-i^{3}\bm{\theta}(\bm{t})\sum_{\vec i}\sum_{p_{1}p_{2}}\sum_{r}p_{{\rm g}_{\vec i}}\zeta_{p_{1}p_{2}}G_{{\rm e}_{p_{2}}{\rm e}_{p_{1}}}(t_{2})\\
 & \times\left\langle \vec{\mu}_{{\rm g}_{\vec i}}^{{\rm e}_{p_{1}}}\vec{\mu}_{{\rm g}_{\vec i}}^{{\rm e}_{p_{1}}}\vec{\mu}_{{\rm e}_{p_{2}}}^{{\rm f}_{r}}\vec{\mu}_{{\rm f}_{r}}^{{\rm e}_{p_{2}}}\right\rangle \emath^{{\rm i}\xi_{{\rm e}_{p_{1}}{\rm g}_{\vec i}}t_{1}-{\rm i}\xi_{{\rm f}_{r}{\rm e}_{p_{2}}}t_{3}+\varphi_{{\rm f}_{r}{\rm e}_{p_{2}}{\rm e}_{p_{2}}{\rm e}_{p_{1}}}^{\ast}(\vec t)}.\nonumber 
\end{align}
Here 
\begin{align}
 & \varphi_{cb{\rm e}_{p_{2}}{\rm e}_{p_{1}}}(\vec t)=-g_{{\rm e}_{p_{1}}{\rm e}_{p_{1}}}(t_{1})-g_{bb}(t_{3})-g_{cc}^{\ast}(t_{3})\nonumber \\
 & -g_{b{\rm e}_{p_{1}}}(t_{1}+t_{2}+t_{3})+g_{b{\rm e}_{p_{1}}}(t_{1}+t_{2})+g_{b{\rm e}_{p_{1}}}(t_{2}+t_{3})\nonumber \\
 & -g_{b{\rm e}_{p_{1}}}(t_{2})+g_{c{\rm e}_{p_{1}}}(t_{1}+t_{2}+t_{3})-g_{c{\rm e}_{p_{1}}}(t_{1}+t_{2})\nonumber \\
 & -g_{c{\rm e}_{p_{1}}}(t_{2}+t_{3})+g_{c{\rm e}_{p_{1}}}(t_{2})+g_{cb}(t_{3})+g_{bc}^{\ast}(t_{3})\nonumber \\
 & +2{\rm i}\Im[g_{c{\rm e}_{p_{2}}}(t_{2}+t_{3})-g_{c{\rm e}_{p_{2}}}(t_{2})-g_{c{\rm e}_{p_{2}}}(t_{3})\nonumber \\
 & +g_{b{\rm e}_{p_{2}}}(t_{2})-g_{b{\rm e}_{p_{2}}}(t_{2}+t_{3})+g_{b{\rm e}_{p_{2}}}(t_{3}).
\end{align}

\end{document}